# The strain on scientific publishing

Mark A. Hanson[1], Pablo Gómez Barreiro[2], Paolo Crosetto[3], Dan Brockington[4,5,6]

**Author correspondence:**
MAH (m.hanson@exeter.ac.uk, ORCID: https://orcid.org/0000-0002-6125-3672)
PGB (p.gomez@kew.org, ORCID: https://orcid.org/0000-0002-3140-3326)
PC (paolo.crosetto@inrae.fr, ORCID: https://orcid.org/0000-0002-9153-0159)
DB (Daniel.Brockington@uab.cat, ORCID: https://orcid.org/0000-0001-5692-0154)

1. *Centre for Ecology and Conservation, Faculty of Environment, Science and Economy, University of Exeter, Penryn, TR10 9FE, United Kingdom*
2. *Royal Botanic Gardens, Kew, Wakehurst, Ardingly, West Sussex RH17 6TN, United Kingdom*
3. *Univ. Grenoble Alpes, INRAE, CNRS, Grenoble INP, GAEL, Grenoble 38000, France*
4. *Institut de Ciència i Tecnologia Ambientals (ICTA), Universitat Autònoma de Barcelona*
5. *ICREA, Pg. Lluís Companys 23, Barcelona, Spain*
6. *El Departament de Dret Privat, Universitat Autònoma de Barcelona*

## 1. Abstract

Scientists are increasingly overwhelmed by the volume of articles being published. Total articles indexed in Scopus and Web of Science have grown exponentially in recent years; in 2022 the article total was ~47% higher than in 2016, which has outpaced the limited growth – if any – in the number of practising scientists. Thus, publication workload per scientist (writing, reviewing, editing) has increased dramatically. We define this problem as "the strain on scientific publishing." To analyse this strain, we present five data-driven metrics showing publisher growth, processing times, and citation behaviours. We draw these data from web scrapes, requests for data from publishers, and material that is freely available through publisher websites. Our findings are based on millions of papers produced by leading academic publishers. We find specific groups have disproportionately grown in their articles published per year, contributing to this strain. Some publishers enabled this growth by adopting a strategy of hosting "special issues," which publish articles with reduced turnaround times. Given pressures on researchers to "publish or perish" to be competitive for funding applications, this strain was likely amplified by these offers to publish more articles. We also observed widespread year-over-year inflation of journal impact factors coinciding with this strain, which risks confusing quality signals. Such exponential growth cannot be sustained. The metrics we define here should enable this evolving conversation to reach actionable solutions to address the strain on scientific publishing.

*Keywords: Open access, Web of Science, Scopus, Impact Factor*

## 2. Introduction

Academic publishing has a problem. The last few years have seen an exponential growth in the number of peer-reviewed journal articles, which has not been matched by the training of new researchers who can vet those articles (Fig. 1A). Editors are reporting difficulties in recruiting qualified peer reviewers (Fox et al., 2017; Peterson et al., 2022), and scientists are overwhelmed by the volume of new articles being published (Parolo et al., 2015; Severin & Chataway, 2021). We will call this problem "the strain on scientific publishing".

Part of this growth may come from inclusivity initiatives or investment in the Global South, which make publishing accessible to more researchers (Maher et al., 2020; Nakamura et al., 2023). Parallel efforts have also appeared in recent years to combat systemic biases in scientific publishing (Langin, 2023; Liu et al., 2019; Meijaard et al., 2015), including positive-result bias (Mlinarić et al., 2017). To the extent that the growth comes from such initiatives, it is welcome and should be accommodated.

However, growth will become strain if it compromises the ability of scientists to be rigorous when vetting information (Hofseth, 2018). If scientific rigour is allowed to slip, it devalues the term "science" (Sarewitz, 2016). Recent controversies already demonstrate this threat, as research paper mills operating within publishing groups have caused mass article retractions (Abalkina, 2023; Bishop, 2023; Candal-Pedreira et al., 2022; Else & Van Noorden, 2021), alongside renewed calls to address so-called "predatory publishing" (Grudniewicz et al., 2019).

To understand the forces that contribute to this strain, we first present a simple schematic to describe scientific publishing. We then specifically analyse publishers, as their infrastructures regulate the rate at which growth in published articles can occur. To do this, we identify five key metrics that help us to understand the constitution and origins of this strain: growth in total articles and special issues, differences in article turnaround times or rejection rates, and a new metric based on citation behaviour that we call "impact inflation."

These metrics should be viewed in light of publisher business models. First, there is the more classic subscription-based model generating revenue from readers. Second, there is the "gold open access" model, which generates revenue through article processing charges that authors pay. In both cases publishers can act either as for-profit or not-for-profit organisations. We therefore consider if aspects of either of these business models are contributing to the strain.

By performing a comparative analysis combining multiple metrics, we find strain is not strictly tied to any one publisher business model, although some behaviours are associated with specific gold open access publishers. We argue that existing efforts to address this strain are insufficient. We highlight specific areas needing transparency, and actions that publishers, researchers, and funders can take to respond to this strain. Our study provides essential data to inform the existing conversation on academic publishing practices.

## 3. The triangle of scientific publishing: a conceptual framework

The strain on scientific publishing is the result of interactions between three sets of players: publishers, funders, and researchers.

Publishers want to publish as many papers as possible, subject to a quality constraint, which varies from brand to brand. This is because publishers advertise total articles published as a desirable quality that augments their brand (File S1), as bargaining chips to negotiate subscription fees (Khelfaoui & Gingras, 2022; Shu et al., 2018), and, when gold open access is involved, publishers earn revenue in direct proportion to total articles published (Butler et al., 2023). Journal brands (e.g. "Nature") can also be used to market other services, or even other journals (Khelfaoui & Gingras, 2022). Publications themselves give researchers a "stamp of quality" that researchers use to advance their own goals. The quality of a stamp is often determined by journal-level prestige metrics, such as the Clarivate journal Impact Factor (IF), or Scopus Scimago Journal Rank (SJR) (Garfield, 2006; Guerrero-Bote & Moya-Anegón, 2012), and ultimately by association with the quality of published papers. Publishers compete amongst each other to attract the most and/or the best papers.

Funders (e.g. universities, funding agencies) use "stamps" from the science publication market as measures of quality to guide their decisions on whom to hire and fund. Ultimately, money from funders supports the whole market, and funders want cost-effective, stable, and informative signals to help guide their decisions. Some funders are experimenting with ways to reward researchers independent from publisher stamps (e.g. narrative CVs (DORA, 2022), though others use journal stamps as signals to determine researcher promotion or salary (e.g. (Quan et al., 2017)).

Finally, due to "publish or perish" pressures, researchers are incentivized to publish as many papers in prestigious journals as possible (Quan et al., 2017; Sarewitz, 2016). If publications are among the main outputs used by funders to gauge researcher productivity then researchers must be productive, which becomes synonymous with 'publish', to secure employment, promotion, and funding. Researchers are also the backbone of science, acting as authors that generate articles, but also as referees and editors during peer review for publishers and funders, almost always for free. As a result, they help influence the administering of publisher stamps of quality. More altruistically, they help ensure the quality of science in their field.

The incentives for publishers and researchers to increase their output drive growth. This is not problematic per se, but there will be a trade-off between the volume of work being produced and its quality. The difficulty is that "quality" is hard to define (Garfield, 2006; Guerrero-Bote & Moya-Anegón, 2012; Thelwall et al., 2023), and some metrics are at risk of abuse per Goodhart's law: "*when a measure becomes a target, it ceases to be a good measure*" (Fire & Guestrin, 2019). For instance, having many citations may indicate an author, article, or journal, is having an impact. But, citations can be gamed through self-citing or coordinated "citation cartels" (Abalkina, 2023; Bishop, 2023; Fong et al., 2023).

Collectively, the push and pull by the motivations of these players defines the sum product of the scientific publishing industry.

## 4. Materials and Methods

### Description of publisher data

We produced five metrics of publisher practice that describe the total volume of material being published, or that affect the quality of publisher "stamps". We focused our analyses on the last decade of publication growth, with special attention paid to the period 2016-2022, as pre-2016, some data types were less available. We used the Scopus database (via Scimago (Scimago, 2023)) filtered for journals indexed in both Scopus and Web of Science. Our focus on journals indexed by both Scopus and Web of Science describes a more conservative measure of journal data compared to other possible datasets (e.g. Dimensions, OpenAlex), but benefits from focussing only on journals that pass stricter indexing criteria. We further assembled journal/article data by scraping information in the public domain from web pages, and/or following direct requests to publishers (Table 1). These metrics are:

- Total articles indexed in both Scopus and Web of Science
- Share of articles appearing in special issues
- Article turnaround times from submission to acceptance
- Journal rejection rates as defined by publishers
- A new metric we call "impact inflation," informed by journal citation behaviours

| Publisher | Web scraped journals included | Articles with turnaround times[†] | Total journals (Scimago) | Total articles (Scimago)[†] |
|---|---|---|---|---|
| BMC* | 289 | 344501 | 213 | 241493 |
| Elsevier | 376 | 531580 | 1579 | 2988422 |
| Frontiers | 44 | 291017 | 49 | 329370 |
| Hindawi* | 220 | 226612 | 161 | 155396 |
| MDPI | 98 | 838448 | 152 | 840518 |
| Nature* | 144 | 360855 | 111 | 346845 |
| PLOS | 12 | 243398 | 7 | 148404 |
| Springer | 1259 | 1371405 | 1589 | 1378386 |
| Taylor & Francis | 1063 | 512438 | 1160 | 525029 |
| Wiley | 1257 | 890174 | 1467 | 1450487 |

[†] Time period 2016-2022

**Table 1: summary of web scraped data informing share of special issue articles and turnaround times.** For some publishers, the number of web scraped journals or articles with turnaround time data exceeds the totals from our Scimago dataset (noted with *). This is because, in the web-scraped dataset, we included all journals by a given publisher, even if they were not indexed, or indexed by only one of Scopus or Web of Science.

## Data collection

*Publisher and journal-level data: publisher selection*

Due to technical limits of web-scraping, we focused our analyses of special issue proportions, turnaround times, and rejection rates, on only a subset of publishers and articles. We chose publishers according to their size and importance, and to get a sufficient variety of business models. We included Elsevier, Wiley-Blackwell (Wiley), Springer, Nature and Taylor & Francis, which are among the largest long-established academic publishers, and which use a mixture of subscription and gold open access licences. Multidisciplinary Publishing Institute (MDPI), Frontiers Media (Frontiers) and Hindawi are younger for-profit gold open access publishers well known for their business model that publishes many articles through special issues. In contrast, BMC is a for-profit gold open access publisher that operates hundreds of journals but does not publish many special issues by comparison. Finally, PLOS is a not-for-profit gold open access publisher, and among the largest publishers in terms of articles per journal per year.

Previous analyses have grouped journals under the umbrella of their parent company ownership (Butler et al., 2023; Larivière et al., 2015). Here we have aggregated publisher labels according to their corporate branding in Scimago: for instance, Elsevier BV, Elsevier Ltd and similar were aggregated as "Elsevier", or Springer GmbH & Co, Springer International Publishing AG as "Springer." However, we have not grouped subsidiary brands under the umbrella of their parent company (e.g. Nature and BMC journals are owned by the conglomerate "Springer-Nature"). We did this because the behaviour of these subsidiary groups is different from the parent company (sometimes radically so). Certain subsidiary publisher brands further define meaningful groups for comparison (e.g. BMC, Hindawi). The status of publisher brands has also changed over the years: a majority share in Frontiers was once owned by Nature Publishing Group (Enserink, 2015), and Hindawi was purchased by Wiley in 2021, who in 2023 decided to "sunset" the brand following paper mill scandals (Bishop, 2023). If we merged these brands with their parent companies we would obscure and confuse relevant data comparisons. For instance, Hindawi would merge into Wiley in 2022 causing dropout of Hindawi from the 2022 dataset and a spike in Wiley's total articles and use of Special Issue publishing by virtue of a merger, and not through a change in behaviour of pre-existing Wiley journals.

*Publisher and journal-level data: data curation*

Total articles published per year come from a subset of the Scopus database obtained from Scimago (Scimago, 2023). Historical data (1999 to 2022) for total number of articles, total citations per document over 2 years, the Scimago Journal Rank (SJR) metric, and total references per document were obtained from the Scimago web portal (https://www.scimagojr.com/journalrank.php). Scimago yearly data were downloaded with the "only WoS journals" filter applied to ensure the journals we include here were indexed by both Scopus and Web of Science (Clarivate). Within-journal self-citation rate was obtained from Scimago via web scraping.

Historical Impact Factor data (2012-2022) for a range of publishers (16,174 journals across BMC, Cambridge University Press, Elsevier, Emerald Publishing Ltd., Frontiers, Hindawi, Lippincott, MDPI, Springer, Nature, Oxford University Press, PLOS, Sage, Taylor & Francis, and Wiley-Blackwell) were downloaded from Clarivate. Due to the download limit of 600 journals per publisher, these IFs represent only a subset of all IFs available.

Rejection rates were collected from publishers in a variety of ways: 1) obtained from online available publisher reports (Frontiers: https://progressreport.frontiersin.org/peer-review)), 2) given by publishers upon request (PLOS, Taylor & Francis) and 3) web scraping of publicly-available data extracted from the journal or company websites (MDPI, Hindawi, Elsevier via https://journalinsights.elsevier.com/). Frontiers rejection rate data lack journal-level resolution, and are instead the aggregate of the whole publisher corpus per year.

Data on publisher values, goals, motivations and aspirations (File S1) were obtained from annual reports, press releases and reports to shareholders

*Article-level data*

Several methods were used to obtain article submission and acceptance times which were used to calculate turnaround times, here defined as the time taken from submission to formal acceptance. Turnaround times advertised by publishers often refer to the time taken until first decision (either accept or reject), which is affected by the rate of submissions and the rate of rejection, two metrics that are not readily observed. By focusing on turnaround times from submission to acceptance, we strictly assess the time spent considering papers deemed worth publishing, which better reflects the rigour of the peer review process for a given publisher, is a comparable metric across publishers, and is readily available on article's websites. PLOS, Hindawi and Wiley´s turnaround times were extracted directly from their corpus. The latter was shared with the authors by Wiley upon request, while the corpuses of PLOS ([https://plos.org/text-and-data-mining/](https://plos.org/text-and-data-mining/)) and Hindawi ([https://www.hindawi.com/hindawi-xml-corpus/](https://www.hindawi.com/hindawi-xml-corpus/)) are available online. For other turnaround times, we used web scraping, extracting the publicly reported date of submission and date of acceptance from article web pages.

Annotation of whether articles were contained in "special issues" required some degree of manual data curation. Broadly, if publishers defined articles as belonging to a "special issue" or other article collection heading (e.g. "Theme Issues", "Collections" or "Topics"), we considered those articles to be special issue articles. Per definitions from COPE (COPE, 2023; Wager, 2012), these articles are often handled by guest editors, but special issues can also be handled by in-house editors. Special issues also vary in format and editorial practice across publishers. As such, we have used an inclusive approach to describing article collections as special issues, ensuring we do not filter out article collections for vague reasons.

*Web scraping*

Web scraping scripts considered a publisher's website formatting and article metadata formatting, and so required manual curation to tailor them to each publisher. BMC, Frontiers, MDPI, Nature and Springer data were obtained via web scraping of individual articles and collecting data in "article information"-type sections. Taylor & Francis turnaround times were obtained via CrossRef (*CrossRef*, 2023) by filtering all available ISSNs from Scimago. To obtain Elsevier turnaround times we first extracted all Elsevier related ISSNs from Scimago, queried these in CrossRef to obtain a list of DOIs, and then web scraped the data from those articles. We also collected information on whether Elsevier articles were part of special issues during our web-scraping. However, the resulting data were unusually spotty and incomplete: for instance, we had journals with a total of only one article with data on special issue status, which would falsely suggest that 100% of articles in that journal were special issue articles. Ultimately, we did not include Elsevier in our analysis of special issue articles.

While web scraping had to be tailored to each publisher website, the core strategy was consistent across web-scraped publishers. In summary, URLs of published articles were obtained in order to be able to download their HTML or XML code and inspect nodes containing information relevant to editorial times and special issues. Extracted data was then formatted to be consistent and allow publisher-to-publisher comparisons.

*Global researcher statistics*

Total PhD graduate numbers were obtained from the Organisation for Economic Co-operation and Development (OECD, 37 countries at the time of data collection: https://stats.oecd.org, data up to 2020). Other sources were consulted to complement OECD data with PhD graduate data for China and India (NSF, 2022; Zwetsloot et al., 2021) but data for these countries only existed up to 2019 and were not collected using the same parameters as OECD data. We therefore present PhD data on OECD countries in the main text, and confirmed and estimated data trends beyond OECD data in the supplementary material.

We further complemented PhD graduate data with measures of researchers-per-million (full time equivalent) from the February 2023 release of the UNESCO Science, Technology, and Innovation dataset (http://data.uis.unesco.org, "9.5.2 Researchers per million inhabitants") and projected these data to 2022 using a linear regression model.

*Rejection rates*

We defined rejection rate as a function of accepted, rejected, and total submissions, depending on the data that were available for each publisher. Of note, rejection rates are affected by a number of cryptic factors that make direct comparisons complicated. For example, journals may differ in their culture for how long revisions are permitted before a resubmission is classified as a new article.

For Frontiers, we use *1 - (accepted articles / total submissions)* to be consistent with data available for other publishers. We defined total submissions for MDPI as the sum of all accepted and rejected articles and calculated rejection rates accordingly. Hindawi report their rejection rates publicly on journal pages as "acceptance rate," although the underlying calculation method is not given.

While we could not standardise the methodology used to calculate rejection rates across publishers, we make the assumption that publishers have at least maintained a consistent methodology internally across years. For this reason, while comparing raw rejection rates comes with many caveats, comparing the direction of change itself in rejection rates within groups should be relatively robust to allow comparisons of trends between groups.

*Impact Inflation*

Clarivate's Impact Factor (IF) is calculated as the mean total citations per article in articles published within the last two years. The Scimago Journal Rank (SJR) is more complex, using a citation network approach that places a higher value on citations between journals of the same general field and limits the contribution to the SJR made by journals through self-citation or citation cartel behaviours. Full details of the SJR metric are given in Guerrero-Boté and Moya-Anegón (Guerrero-Bote & Moya-Anegón, 2012). Key differences between the two metrics are summarised in Table 2. Importantly, SJR limits the reward in prestige/rank that journals receive from individual sources. Thus, while high rates of self-citation or circular citation are rewarded indefinitely by IF, their contribution to a journal's SJR is limited. The ratio of IF/SJR can therefore reveal journals whose total citations come from disproportionately few citing journals. We call this ratio "Impact Inflation," which applies to both IF/SJR, or the equivalent calculation using Scimago's IF equivalent "citations per document (Cites-per-doc /SJR)."

**Table 2: Methodological differences between SJR and Impact Factor.** Table content adapted from Guerrero-Boté and Moya-Anegón (Guerrero-Bote & Moya-Anegón, 2012).

|  | **SJR** | **Impact Factor** |
| --- | --- | --- |
| **Database** | Scopus | Web of Science |
| **Citation time frame** | 3 years | 2 years |
| **Self-citation contribution** | Limited | Unlimited |
| **Field-weighted citation** | Weighted | Unweighted |
| **Size normalisation** | Citable document rate | Citable documents |
| **Citation networks considered?** | Yes | No |

We describe publisher characteristics using journal-level values (within journal self-citation rate, IF/Cites-per-doc). Notably, the age of journals is tied to article output, as newer journals publish fewer articles, but can grow to publish thousands of articles annually in later years. New journals are also likely to have lower self-citation rates as, at their inception, they have few articles available to self-cite. Small journals may also have different self-citation behaviours from larger journals if, for example, they cite their own previous papers for things like geographic reasons (e.g. the Canadian Journal of Fisheries and Aquatic Sciences). Therefore, we analyse journal data with journal size considered, and filtered for established journals whose articles receive a total of at least 1000 annual citations for self-citation analyses. This was especially important for comparisons at the publisher level, as some publishers have increased their number of journals substantially in recent years (Fig. S1), meaning a large fraction of their journals are relatively young and less characteristic of the publisher's trends according to their better-established journals.

High resolution versions of all the figures can be found at:

https://figshare.com/articles/figure/The_strain_on_scientific_publishing_figures_/24203790

### Data Availability

Due to copyright concerns over our web scraping of information in the public domain, we were legally advised not to release our data and scripts publicly. In pursuit of principles of Open Science, we have shared what we can in the supplementary materials (including README files explaining to the reader how to assemble our data, File S2). We have further developed a web app (via R shinyapp) hosted on the web page https://the-strain-on-scientific-publishing.github.io/website/ that allows users to interrogate our data for specific trends. The app includes publishers not highlighted in the main text of this article. For more details see: https://pagoba.shinyapps.io/strain_explorer/.

## 5. Results

### A few publishers disproportionately contribute to total article growth

There were ~897k more indexed articles per year in 2022 (~2.82m articles) compared to 2016 (~1.92m articles) (Fig. 1A), a year-on-year exponential growth of ~5.6% over this time period. To understand the source of this substantial growth, we first divided article output across publishers per Scopus publisher labels (Fig. 1B). The five largest publishers by total article output include Elsevier, MDPI, Wiley, Springer, and Frontiers respectively. However, in terms of strain added since 2016, their rank order changes: journals from MDPI (~27%), Elsevier (~16%), Frontiers (~11%), Springer (~9.5%), and Wiley (~7.0%) have contributed >70% of the increase in articles per year. Elsevier and Springer own a huge proportion of total journals, a number that has also increased over the past decade (Fig. S1). As such, we normalised article output per journal to decouple the immensity of groups like Elsevier and Springer from the growth of articles itself. While Elsevier has increased article outputs per journal slightly, other groups such as MDPI and Frontiers, have become disproportionately high producers of published articles per journal (Fig. 1C).

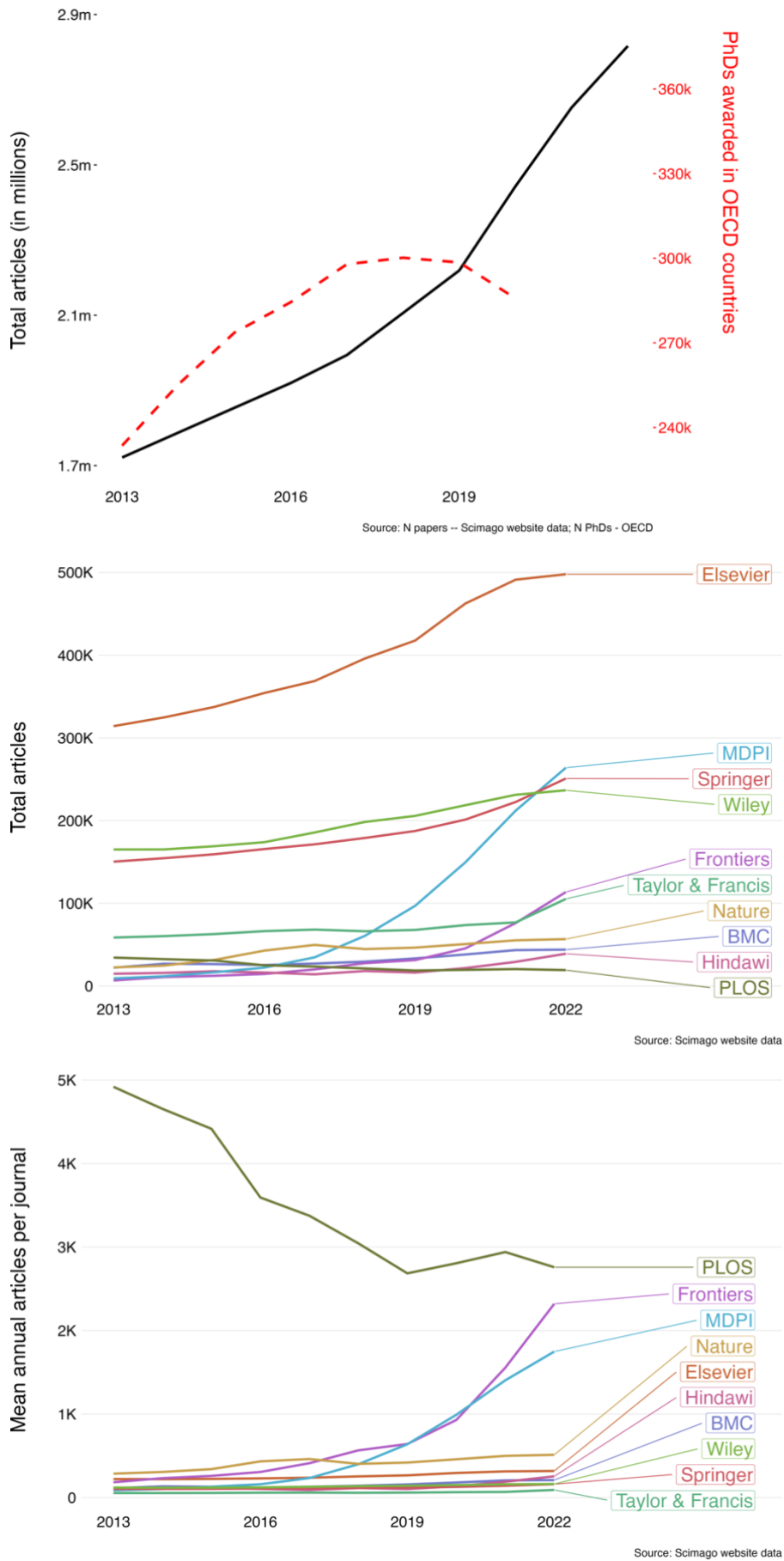

**Figure 1: Total article output is increasing.** A) Total articles being published per year has increased exponentially, while PhDs being awarded have not kept up. This remains true with addition of non-OECD countries, or when using global total employed researcher-hours instead of PhD graduates as a proxy for active researchers (Fig. S2). B-C) Total articles per year by publisher (B), or per journal per year by publisher (C). Also see growth in journals per publisher (Fig. S1) and by size class (Fig. S3).

Taken together, groups like Elsevier and Springer have quantitatively increased total article output by distributing articles across an increasing number of journals. Meanwhile groups like MDPI and Frontiers have been exponentially increasing the number of publications handled by a much smaller pool of journals. These publishers reflect two different mechanisms that have promoted the exponential increase in total articles published over the last few years.

**Growth in articles published through "special issues"**

"Special issues" are distinct from standard articles because they are invited by journals or editors, rather than submitted independently by authors. They also commonly delegate responsibilities to guest editors, whereas editors for regular issue articles are members of their journal's editorial board, expected to have more experience with their journal's specific expectations. In recent years, certain publishers have adopted this business model as a route to publish the majority of their articles (Fig. 2). This behaviour encourages researchers to generate articles specifically for special issues, raising concerns that publishers could abuse this model for profit (Butler et al., 2023; Ioannidis et al., 2023; Larivière et al., 2015). Here we describe this growth in special issues for eight publishers for which we could collect data.

Between 2016 and 2022, the proportion of special issue articles grew drastically for Hindawi, Frontiers, and MDPI (Fig. S4, S5). These publishers depend on article processing charges for their revenues, which are paid by authors to secure gold open access licences. But this special issue growth is not a necessary feature of open access publishing as similar changes were not seen in other gold open access publishers (i.e. BMC, PLOS). Publishers using both subscription and open access approaches (Nature, Springer, Wiley) also tended to publish small proportions of special issues.

These data show that the strain generated by special issues is not a direct consequence of the rise of open access publishing per se, or associated article processing charges. Instead, the dominance of special issues in a publisher's business model is publisher-dependent.

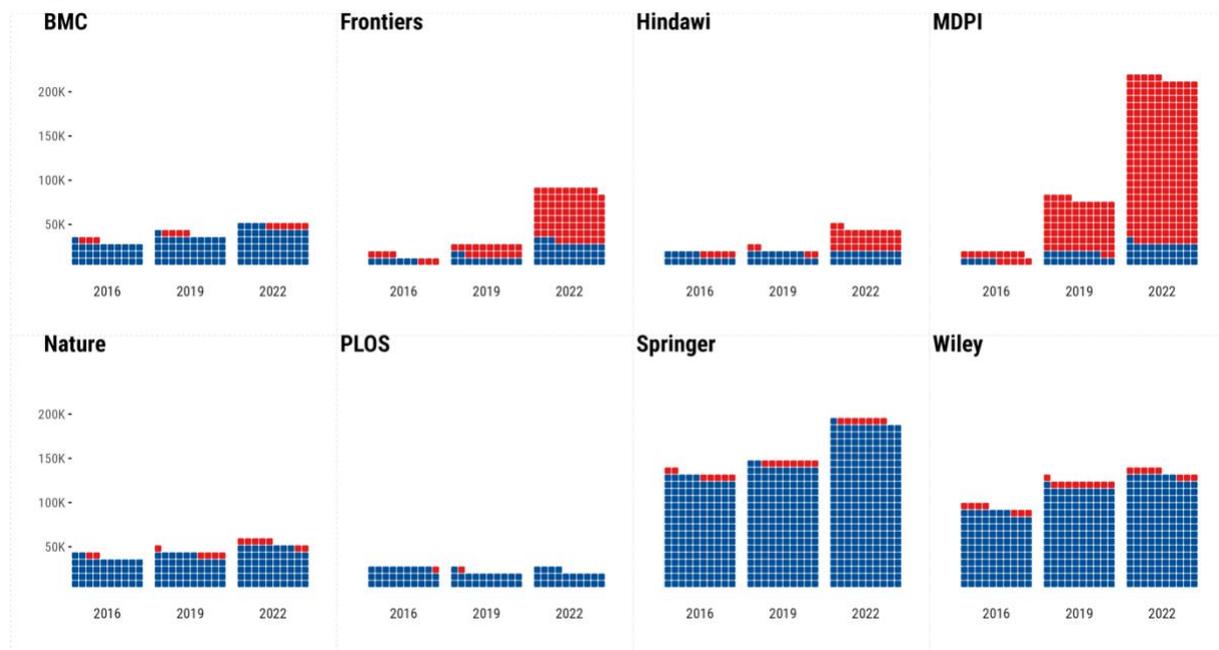

**Number of papers published in regular vs special issues, 2016-22**
One square = 800 articles

Source: data scraped from the publisher's website

Note: Special issues are called Collections at PLOS and Topics at Frontiers. For MDPI Collections, Sections and Topics not shown.

**Figure 2: rise of the special issue model of publishing.** Normal articles (blue) and special issue articles (red) over time. Frontiers, Hindawi, and especially MDPI publish a majority of their articles through special issues, including an increase in recent years alongside growth seen in Fig. 1 (detailed further in Fig. S4, S5). These data reflect only a fraction of total articles shown in Fig. 1, limited due to sampling methodology (Table 1). Elsevier are excluded because of problems of data availability.

**Decreasing mean, increasing homogeneity of turnaround times**

The time spent in processing an article is colloquially known as its "turnaround time." Article turnaround times rely on the rapidity of the publisher to provide an editorial decision, and the level of care that researchers take in making edits during revisions before resubmitting the article for consideration. Here we analyse article turnaround times, defined as the time taken from first submission to editorial acceptance. We use this timeframe because the time required for revisions can take weeks to months depending on field of research and the magnitude or type of revisions required; for instance, minor text changes could be made quickly, but essential experiments requested by reviewers could take months to complete. As a result, turnaround times within journals should be highly heterogenous if each article is considered and addressed according to its unique needs, as expected of rigorous peer review. Moreover, the average turnaround time is expected to vary from journal to journal according to research field. This is because revision requirements vary by research field, for example: in biology, additional reviewer-requested experiments might take just a few weeks to complete using a fruit fly study system, while experiments with mice might take months

We analysed turnaround times between 2016 and 2022 for publications where data were available. We found that average turnaround times vary markedly across publishers. Like others (MDPI, 2021; Oviedo-García, 2021), we found that MDPI had an average turnaround time of ~37 days from first submission to acceptance in 2022, a level they have held at since ~2018. This turnaround time is far lower than comparable publishers like Frontiers (72 days) and Hindawi (83 days), which also saw a decline in mean turnaround time between 2020 and 2022. On the other hand, other publishers in our dataset had turnaround times of >130 days, and if anything, their turnaround times increased slightly between 2016-2022 (Fig. 3A).

The publishers decreasing their turnaround times also show declining variances. Turnaround times for Hindawi, Frontiers, and especially MDPI are becoming increasingly homogenous (Fig. 3B and S6): articles, regardless of initial quality or field of research, and despite the expectation of heterogeneity, are all accepted in an increasingly similar timeframe.

The decrease in mean turnaround times (Fig. 3A) also aligns with inflection points for the exponential growth of articles published as part of special issues in Hindawi (2020), Frontiers (2019), and MDPI (2016) (see Fig. S4). We therefore asked if special issue articles are processed more rapidly than normal articles in general. For most publishers, this was indeed the case, even independent of proportions of normal and special issue articles (Fig. S7). In the case of Hindawi, a previous study also noted special issue articles were more likely to involve paper mill activity, suggesting both lower article quality and lower editorial rigour (Bishop, 2023).

In summary turnaround times differ by publisher, associated with use of the special issue publishing model. Variance in turnaround times also decreases for publishers alongside adoption of the special issue model. These results suggest that special issue articles are typically accepted more rapidly and in more homogenous timeframes than normal articles, which, to our knowledge, has never been formally described.

A

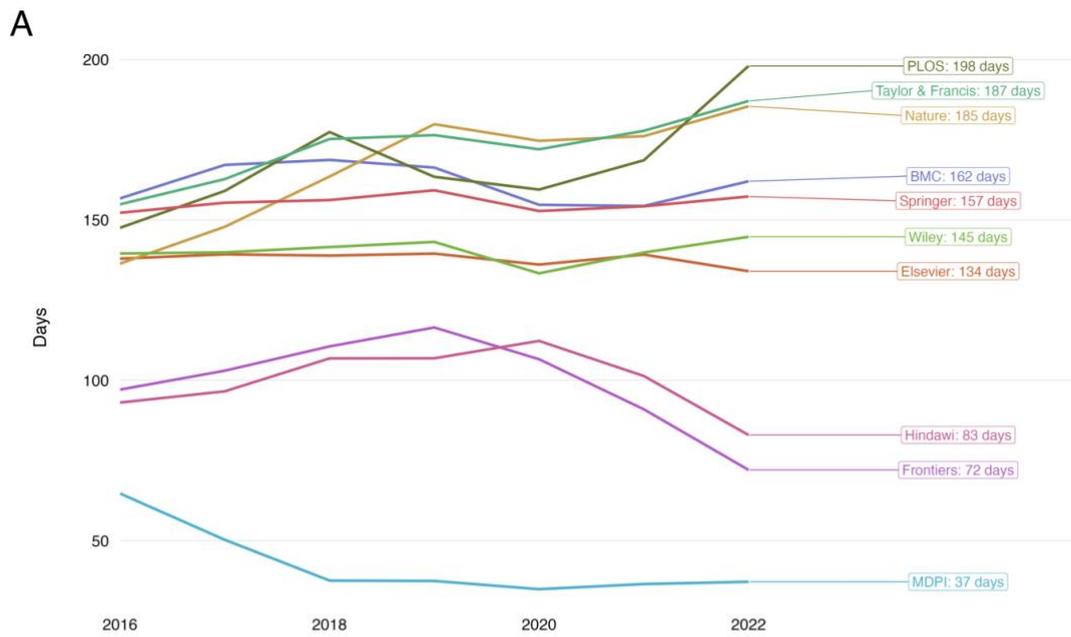

B

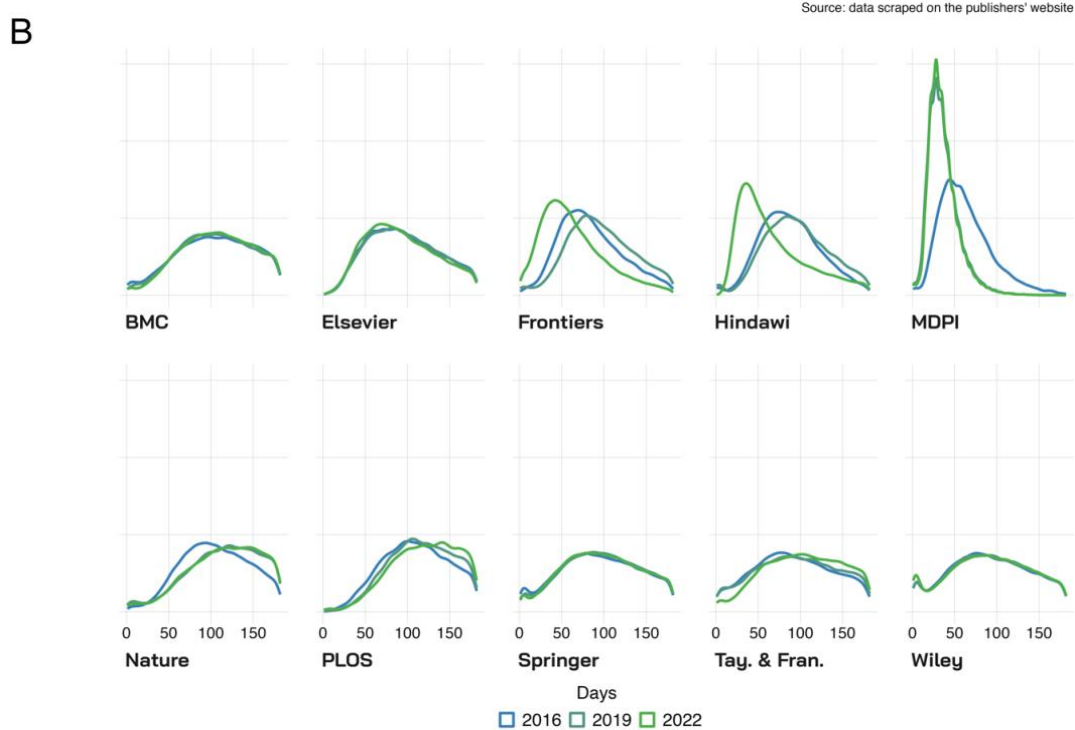

**Figure 3: Article turnaround times.** A) Evolution of mean turnaround times by publisher. Only articles with turnaround times between 1 day and 2 years were included. This filter was applied to remove data anomalies such as immediate acceptance or missing values that default to Jan 1st 1970 (the "Unix epoch"). B) Article turnaround time distribution curves from 2016-2022, focused on the first six months to better show trends. While most publishers have a right-skewed curve, the three publishers highlighted previously for increased special issue use have a left-skewed curve that only became more extreme over time. These data reflect only a fraction of total articles shown in Fig. 1, limited due to sampling methodology (see Table 1). Tay. & Fran. = Taylor & Francis. Full descriptive statistics in Table S1.

## Journal rejection rates and trends are publisher-specific

If a publisher lowers its article rejection rates, all else being equal, this will lead to more articles being published. Such changes to rejection rate might also mean more lower-quality articles are being published. Peer review is the principal method of quality control that defines science (Grainger, 2007), and so publishing more articles with lower quality may add to strain and detract from the meaning and authority of the scientific process. The relationship between rejection and quality is complex: high or low rejection rates may stem from editorial oversight over scope, or perceived impact. Publishers also define what it means to be "rejected," creating caveats to comparing raw numbers across publishers.

Rejection rate data are rarely made public, and only a minority of publishers provide these data routinely or shared rejection rates upon request. Using the rejection rate data we could collect, we estimated rejection rates per publisher and asked if they: 1) change with growth in articles, 2) correlate with journal size, 3) correlate with journal impact, 4) depend on the publisher, or 5) predict a journal's proportion of species issue articles.

We found no clear trend between the evolution of rejection rates and publisher growth (Fig. 4A). Focussing on younger journals (≤10 years, ensuring fair comparisons) we found no relationship between journal size and reported or calculated 2022 rejection rates (Fig. 4B). Finally, citations per document (similar to Clarivate IF) did not correlate with rejection rates (Fig. S8), indicating citations are not a strong predictor. Ultimately, the factor that best predicted rejection rates was the publisher itself: although both Frontiers and MDPI have similar growth in special issue articles (Fig. 2), they show opposite trends in rejection rates over time, and MDPI uniquely showed decreasing rates compared to other publishers (Fig. 4A). Raw rejection rates for MDPI in 2022 were also lower than other publishers. Moreover, Hindawi and MDPI journals with more special issue articles also had lower rejection rates (P = 5.5e-8 and P = .01 respectively, data from 2022, Fig. 4C, Fig. S9), which we could not assess for other publishers.

In summary we found no general associations across publishers between rejection rate and most other metrics we investigated. Over time or among journals of similar age, rejection rate patterns were largely publisher specific. We did, however, recover a trend that within the limited subset of publishers for which we have data, rejection rates decline with increased use of special issue publishing.

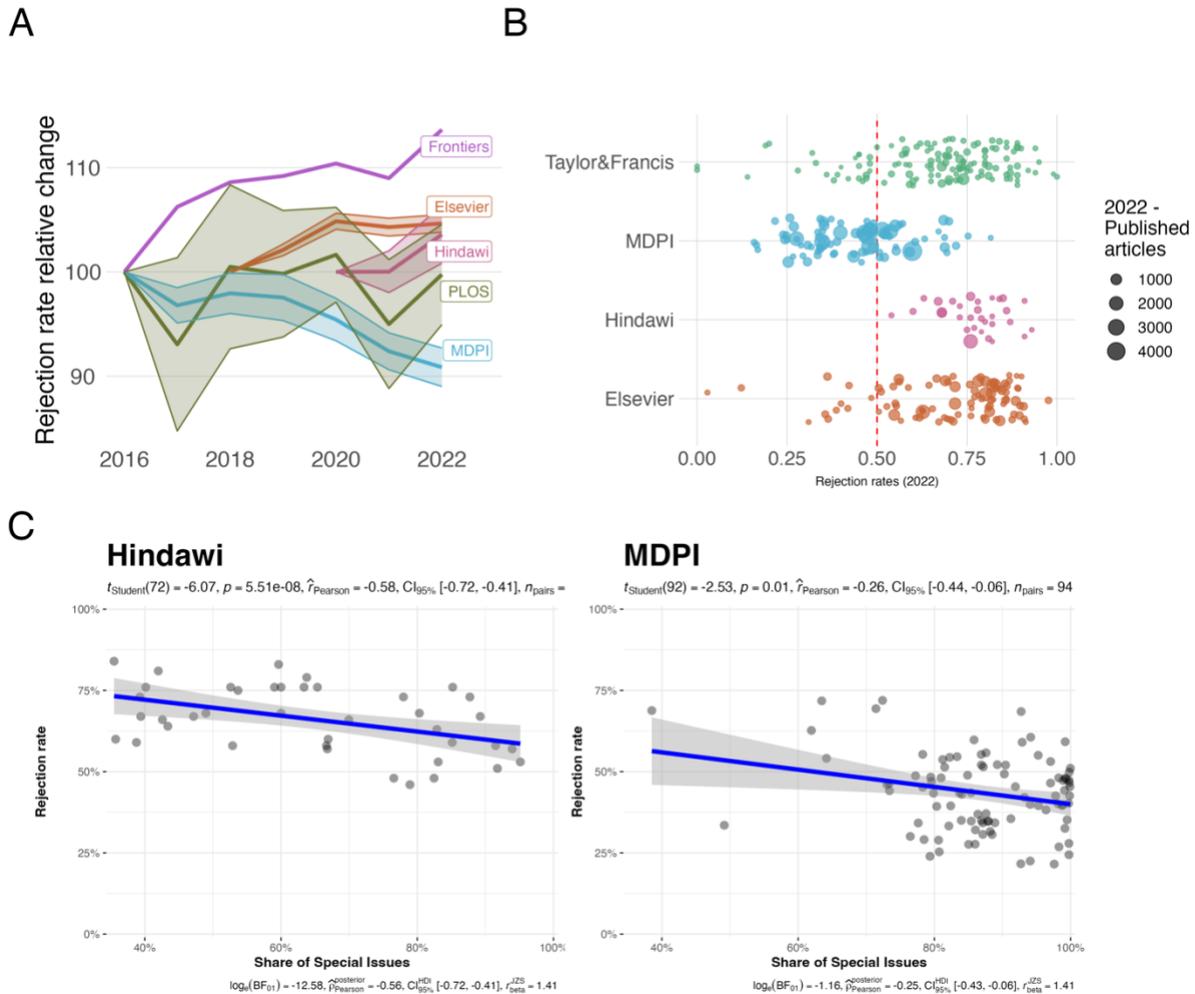

**Figure 4: rejection rates are defined most specifically by publisher.** A) Rejection rates increased, decreased, or experienced little change depending on publisher. We estimated publisher rejection rates from varying available data, so we normalised these data by setting the first year on record as "100." Frontiers data are the aggregate of all Frontiers journals, preventing the plotting of 95% confidence intervals. B) 2022 rejection rates among young journals (<10 years old) differ by publisher, but not journal size. C) For two publishers that we could analyse, there was a significant correlation between 2022 journal rejection rates decreasing and the share of articles published through special issues increasing.

**Disproportionately inflated Impact Factor affects select publishers**

Among the most important metrics of researcher impact and publisher reputation are citations. For journals, the Clarivate 2-year IF reflects the mean citations per article in the two preceding years. Here we found that IF has increased across publishers in recent years (Fig. S10, S11). Explaining part of this IF inflation, we observed an exponential increase in total references per document between 2018-2021 (Fig. S12, and see Neff & Olden, 2010). However, we previously noted that IF is used as a "stamp of quality" by both researchers and publishers to earn prestige, and that IF can be abused by patterns of self-citation. We therefore asked if changes in journal citation behaviour may have contributed to recent inflation of the IF metric

To enable systematic analysis, we used Cites-per-doc from the Scimago database as a proxy of Clarivate IF (Cites-per-doc vs. IF: $R^2$ = 0.77, Fig. S13). We then compared Cites-per-doc to the network-based metric "Scimago Journal Rank" (SJR). Precise details of these metrics are discussed in the supplementary methods (and see Guerrero-Bote & Moya-Anegón, 2012). A key difference between SJR and Cites-per-doc is that SJR has a maximum amount of 'prestige' that can be earned from a single source. As such, if a journal disproportionately cites itself, or if citation cartels coordinate rings of citations between papers hosted by relatively few journals, this will increase the IF and Cites-per-doc of journals hosting such papers, but it will not increase the journal's SJR. We define the ratio of Cites-per-doc to SJR (or IF to SJR) as "impact inflation."

Impact inflation differs dramatically across publishers (Fig. 5A), and has also increased across publishers over the last few years (Fig. S14A). In 2022, impact inflation in MDPI and Hindawi were significantly higher than all other publishers ($P^{adj}$ < .05). Interestingly, Frontiers had low impact inflation comparable to other publishers, despite growth patterns similar to MDPI and Hindawi.

The reason behind MDPI's anomalous impact inflation appears to be straightforward: MDPI journals nearly universally spiked in rates of articles citing other articles from the same journal during the study period (here referred to as "self-citation" rate, Fig. S14B).

There were significant differences in self-citation rates of MDPI journals compared to other publishers (Fig. 5B, $P^{adj}$ < .05 broadly, and MDPI vs. Taylor & Francis, $P^{adj}$ = .13), including comparisons in previous years (S15, $P^{adj\ 2021}$ < .05 broadly, including MDPI vs. Taylor & Francis $P^{adj\ 2021}$ =3e-7). Indeed, beyond within-journal self-citations, in an analysis from 2021, MDPI journals received ~29% of their citations from other MDPI journals (MDPI, 2021), which would be rewarded per citation for IF but less so for SJR. Notably, Hindawi had self-citation rates more comparable to other publishers (Fig. 5B, Fig. S15), despite high impact inflation. In this regard, while Hindawi journals may not directly cite themselves as often, they may receive many citations from a small network of journals, including many citations from MDPI journals (example in Fig. S16), or from paper mills that operated within Hindawi (Bishop, 2023), and may have received biased citations from studies published via the same paper mills outside Hindawi.

A

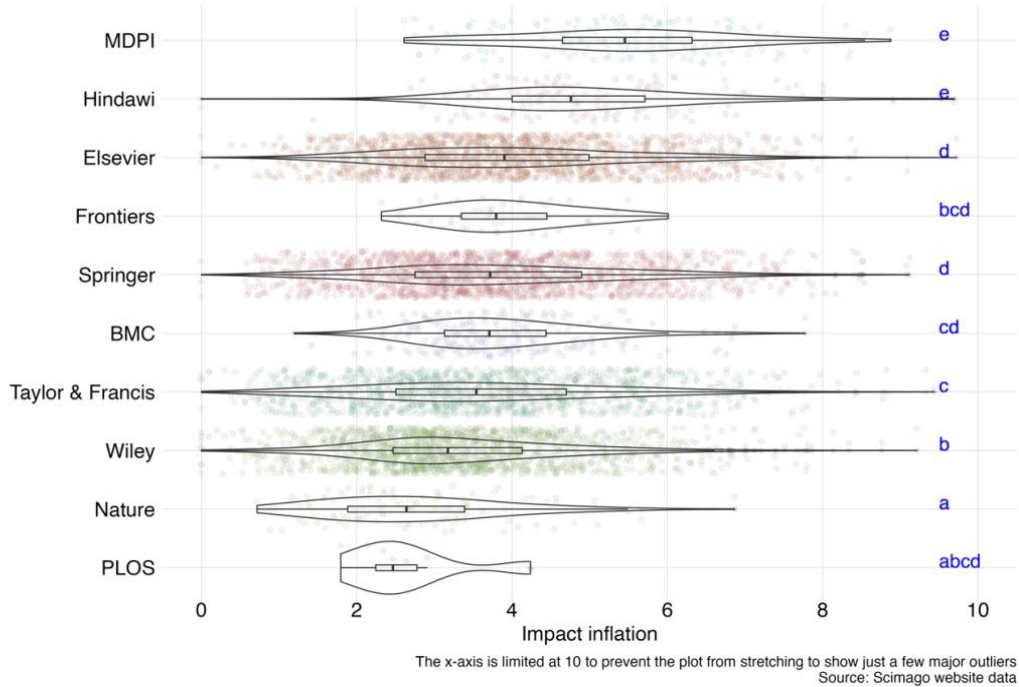

The x-axis is limited at 10 to prevent the plot from stretching to show just a few major outliers
Source: Scimago website data

B

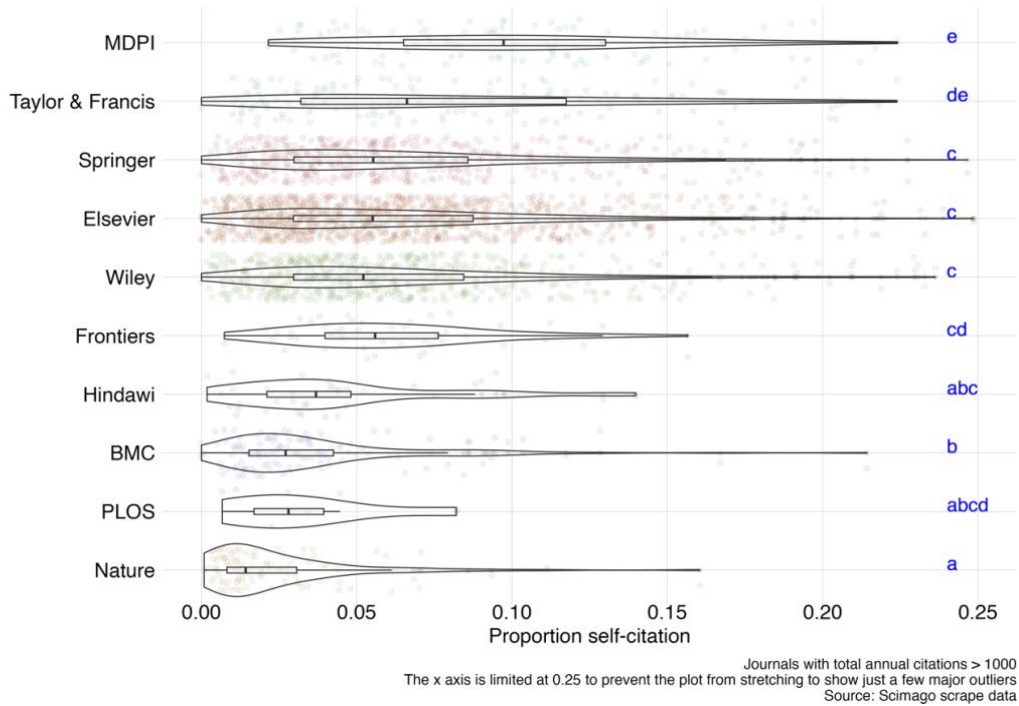

Journals with total annual citations > 1000
The x axis is limited at 0.25 to prevent the plot from stretching to show just a few major outliers
Source: Scimago scrape data

**Figure 5: Changing behaviour of citation metrics revealed by Impact Inflation (for journals with at least 1000 annual citations).** Statistical letter groups reflect significant differences ($P$ < .05) in one-way ANOVA with Tukey HSD. A) MDPI and Hindawi have significantly higher impact inflation compared to all other publishers. Comparisons using samples of Clarivate IFs are shown in Fig. S13. B) MDPI journals have the highest rate of within-journal self-citation among compared publishers, including in previous years (Fig. S14 ,S15). Here we specifically analyse journals receiving at least 1000 citations per year to avoid comparing young or niche journals to larger ones expected to have diverse citation profiles.

In summary, we provide a novel metric, "impact inflation," that uses publicly available data to assess journal citation behaviours. Impact inflation describes how proportionate a journal's total citations are compared to a network-adjusted approach. In the case of MDPI, there was also a high prevalence of within-journal self-citation, consistent with reports by Oviedo-Garcia (2021) and MDPI itself (MDPI, 2021). However high impact inflation and self-citation is not strictly correlated with other metrics we have investigated.

## 6. Discussion

Here we have characterised the strain on scientific publishing, as measured by the exponential rise of indexed articles and the pressure this creates for scientists. The collective addition of nearly one million articles per year over the last 6 years alone costs the research community immensely, both in writing and reviewing time and in fees and article processing charges (Aczel et al., 2021; Shu et al., 2018). Further, given our strict focus on indexed articles, not total articles, our data likely underestimate the true extent of the strain – *the problem is even worse than we describe.*

The strain we characterise is a complicated problem, generated by the interplay of different actors in the publishing market. Funders want to get the best return on their investment, while researchers want to prove they are a good investment. The rise in scientific article output is only possible with the participation of researchers who act as authors, reviewers and editors. Researchers do this because of the "publish or perish" imperative (Grimes et al., 2018), which rewards individual researchers who publish as much as possible, forsaking quality for quantity. On the other hand, publishers host and spur the system's growth in their drive to run a successful business (see File S1). Publishers structure the market, control journal reputation, and as such are focal players – leading to concerns regarding to what extent publisher behaviour is motivated by profit (Butler et al., 2023; Ioannidis et al., 2023; Larivière et al., 2015). Indeed, a previous study found that publishers have used growth of their corpus to negotiate for increases to journal subscription fees, with the result being that the value of money spent per citable article actually declined (Shu et al., 2018). Growth could be welcome if comes from combatting systemic biases in academic publishing. However, if growth does not provide proportionate additional value, it increases strain on the scientific infrastructure.

Considering our metrics in combination (Table 3) also allows us to identify common trends and helps to characterise the role that different publishers play in generating this strain. Across publishers, article growth is the norm, with some groups contributing more than others. Impact factors and impact inflation have both increased universally, exposing the extent to which the publishing system itself has succumbed to Goodhart's law. Nonetheless, the vast majority of growth in total indexed articles has come from just a few publishing houses following two broad models.

**Table 3: Strain indicators from 2016 to 2022.** Data on total articles and impact inflation drawn from the Scimago dataset. Data on special issues, turnaround times, and rejection rates come from web scrapes limited to the publishers shown. Rejection rate change for Elsevier and Hindawi start from 2018 and 2020 respectively. pp = 'percentage points.'

Strain indicators at a glance: 2022 and evolution 2016-22

| | 2022 | | | | | Change 2016-22 | | | | |
|---|---|---|---|---|---|---|---|---|---|---|
| | TOTAL ARTICLES | SHARE SPECIAL ISSUE | TURNAROUND TIME (DAYS) | REJECTION RATE | IMPACT INFLATION | TOTAL ARTICLES | SHARE SPECIAL ISSUE | TURNAROUND TIME (DAYS) | REJECTION RATE | IMPACT INFLATION |
| Overall | 2816k | 38% | 116 | 62% | 3.3 | +47% | +27pp | -23 | -1pp | +1.1 |
| Elsevier | 498k | -- | 134 | 71% | 4.0 | +41% | -- | -4 | +5pp* | +1.5 |
| MDPI | 264k | 88% | 37 | 40% | 5.4 | +1080% | +14pp | -28 | -8pp | +2.2 |
| Springer | 251k | 3% | 157 | -- | 3.9 | +52% | -1pp | +5 | -- | +1.5 |
| Wiley | 237k | 5% | 145 | -- | 3.4 | +36% | -2pp | +5 | -- | +1.2 |
| Frontiers | 114k | 69% | 72 | 48% | 4.0 | +675% | +20pp | -25 | +14pp | +1.8 |
| Taylor & Francis | 105k | -- | 187 | -- | 3.7 | +59% | -- | +32 | -- | +1.5 |
| Nature | 57k | 11% | 185 | -- | 2.8 | +32% | +6pp | +49 | -- | +1 |
| BMC | 44k | 10% | 162 | -- | 3.9 | +73% | +1pp | +5 | -- | +1.5 |
| Hindawi | 39k | 62% | 83 | 74% | 5.0 | +139% | +36pp | -10 | +3pp* | +1.9 |
| PLOS | 19k | 1% | 198 | 59% | 2.6 | -23% | -3pp | +50 | -4pp | +1.1 |

For older publishing houses (e.g. Elsevier, Springer), growth was not driven by major growth across all journals, but by the synergy of mild growth in both total journals and articles per journal in tandem (as discussed in Khelfaoui & Gingras, 2022). Another strategy used by certain for-profit gold open access publishers (e.g. MDPI, Frontiers, Hindawi) consisted of an increased use of special issue articles as a primary means of publishing. This trend was coupled with uniquely reduced turnaround times, and in specific cases, high impact inflation and reduced rejection rates. Despite their stark differences, the amount of strain generated through these two strategies is comparable.

The rich context provided by our metrics also provides unique insights. Ours is the first study, of which we are aware, to document that special issue articles are systematically handled differently from normal submissions: special issues have lower rejection rates, and both lower and seemingly more homogeneous turnaround times. We also highlight the unique view one gets by considering different forms of citation metrics, and develop impact inflation (IF/SJR) as a litmus test for the degree of citation gaming taking place in a journal. Due to paper mill activity infiltrating the academic literature (Abalkina, 2023; Bishop, 2023), there is newfound interest in assessing the citation behaviours of publishing groups (Seeber et al., 2024). Impact Inflation addresses this using publicly available data, making it a valuable open science tool for assessing citation behaviours (in accord with recommendations by (Waltman, 2016)).

Throughout our study MDPI was an outlier in every metric – often by wide margins. MDPI had the largest growth of indexed articles (+1080%) and proportion of special issue articles (88%), shortest turnaround times (37 days), decreasing rejection rates (-8 percentage points), highest impact inflation (5.4), and the highest within-journal mean self-citation rate (9.5%). Ours is not the first study analysing MDPI (Abalkina, 2023; Copiello, 2019; Khoo, 2019; Oviedo-García, 2021), but our broader context highlights the uniqueness of their profile and their contribution to the strain.

Some metrics appear to be principally driven by publisher's policies: rejection rates and turnaround time means and variances are largely independent from any other metric we assayed. This raises questions about the balance between publisher's oversight and scientific editorial independence. This balance is essential to maintain scientific integrity and authority: oversight should be sufficient to ensure rigorous standards, but not so invasive as to override the independence of editors. Understanding how editorial independence is maintained in current publishing environments, though beyond the scope of this paper, is key to maintaining scientific integrity and authority.

Given the importance of scientific publishing, it is unfortunate that the basic data needed to inform an evidence-based discussion are so hard to collect. This discussion on academic publishing would be easier if the metrics we collected were more readily available – we had to web scrape to obtain many pieces of basic information. The availability of our metrics could be encouraged by groups such as the Committee on Publication Ethics (Wager, 2012), which publishes guides on principles of transparency. We would recommend transparency for: proportion of articles published through special issues (or other collection headings), article turnaround times, and rejection rates. Rejection rates in particular would benefit from an authority providing a standardised reporting protocol, which would greatly boost the ability to draw meaningful information from them. While not a metric we analysed, it also seems prudent for publishers to be transparent about revenue and operating costs, given much of the funding that supports the science publishing system comes from taxpayer-funded or non-profit entities. Referees such as Clarivate should also be more transparent; their decisions can have a significant impact on the quality of publisher stamps (see Table S2 and MDPI, 2023), and yet the reasoning behind these decisions is opaque.

Greater transparency will allow us to document the strain on scientific publishing more effectively. However, it will not answer the fundamental question: how should this strain be addressed? Addressing strain could take the form of grassroots efforts (e.g. researcher boycotts) or authority actions (e.g. funder or committee directives, index delistings). Researchers, though, are a disparate group and collective action is hard across multiple disciplines, countries and institutions. Funders can more easily change the publish or perish dynamics for researchers, thus limiting their drive to supply articles.

We recommend funders to review the metrics we define here and adopt policies such as narrative CVs that highlight researchers' best work over total volume (DORA, 2022), which mitigate publish or perish pressures. Indeed, researchers agree that changes to research culture must be principally driven by funders (Wellcome Trust, 2020), whose financial power could also help promote engagement with commendable publishing practices.

Our study shows that regulating behaviours cannot be done at the level of publishing business model. Gold open access, for example, does not necessarily add to strain, as gold open access publishers like PLOS (not-for-profit) and BMC (for-profit) show relatively normal metrics across the board. Rather our findings suggest that addressing strain requires action be taken to address specific publishers and specific behaviours. For instance, collective action by the researcher community, or guidelines from funders or ethics committees, could encourage fewer articles be published through special issues, which our study suggests are held to different standards from normal issues. Indeed, reducing special issue articles would already address the plurality of strain being added. Improved self-governance of publishers by the publishing community, through for example, more vigorous policing of new trends by institutions like COPE, might prevent the sorts of problems we have observed here from occurring.

Here we have characterised the strain on scientific publishing. We hope this analysis helps advance the conversation among publishers, researchers, and funders to reduce this strain and work towards a sustainable publishing infrastructure.

## Acknowledgements

We thank the following publishers for providing data openly, or upon request: MDPI, Hindawi, Frontiers, PLOS, Taylor & Francis, BMC and The Royal Society. We further thank many colleagues and publishers for providing feedback on this manuscript prior to its public release: Matthias Egger, Howard Browman, Kent Anderson, Erik Postma, Yuko Ulrich, Paul Kersey, Gemma Derrick, Odile Hologne, Pierre Dupraz, Navin Ramankutty, and representatives from the publishers MDPI, Frontiers, PLOS, Springer, Wiley, and Taylor & Francis. This work was a labour of love, and was not externally funded.

## Author contributions

Web scraping was performed by PGB and PC, and Scimago data curation by MAH. Global doctorate and global researcher data curation was done by MAH and DB. Data analysis in R was done by MAH, PGB, and PC. Conceptualisation was performed collectively by MAH, PGB, PC, and DB. The initial article draft was written by MAH. All authors contributed to writing and revising to produce the final manuscript.

Acronyms

Impact Factor (IF), Scimago Journal Rank (SJR)

# Supplementary figures and tables

**Total journals by publisher**

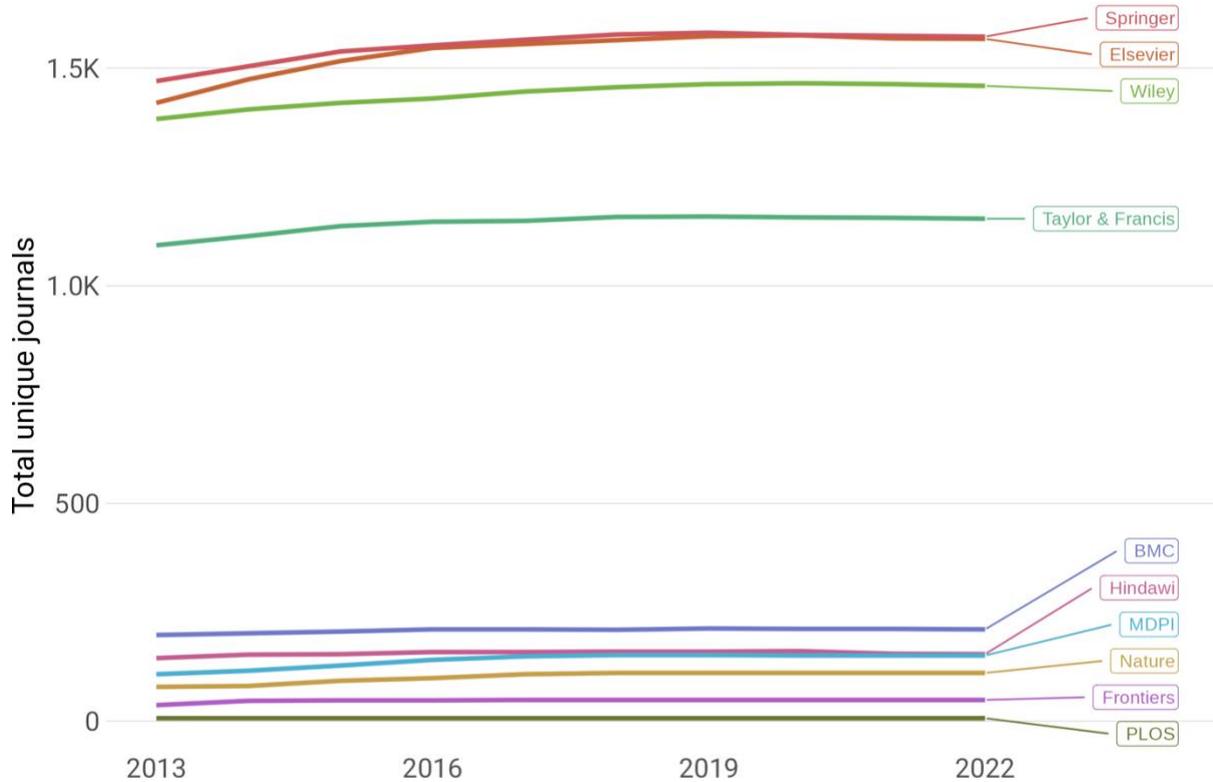

Source: Scimago website data

**Figure S1: growth in total journals by publisher.** Between 2013-2022, Elsevier, Springer, Taylor & Francis, MDPI, and Nature have added to their total journals noticeably. Note: we have only analysed journals indexed in both Scopus and Web of Science, and also collected journals under Publishers according to their licensed Publisher names. Subsidiary publishers under the umbrella of larger publishers are not included in larger publisher totals. For example, both BioMed Central (BMC) and Nature portfolio (Nature) are subsidiaries of Springer Nature (Springer), but host a large number of journals and license under a non-Springer name, and so are treated as separate entities in our study.

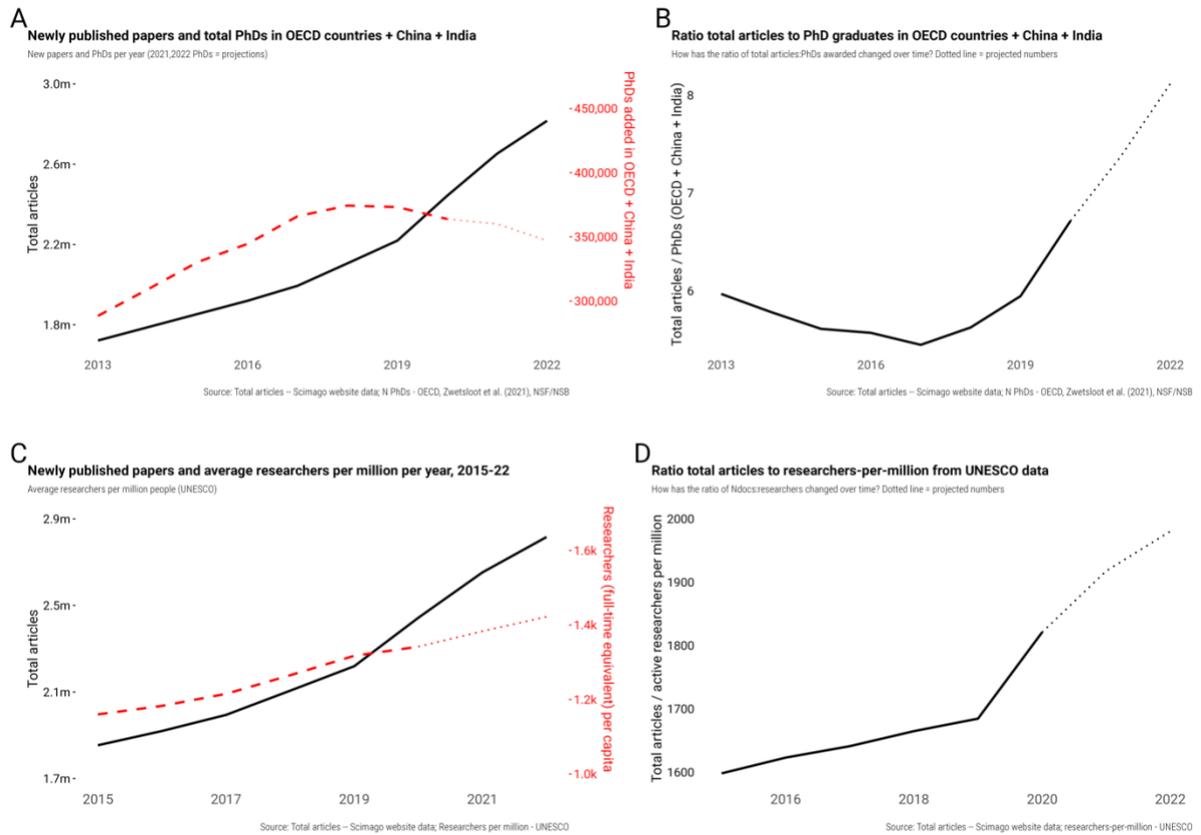

**Figure S2: the growing disparity between total articles per year and active researchers is robust to use of alternate datasets.** Dotted lines indicate estimated trends. A) OECD data complemented with total STEM PhD graduates from India and China (dashed red line) does not alter the pattern of an overall decline in recent years (Fig. 1A). B) The ratio of total articles to total PhD graduates has gone up substantially since 2019. C) UNESCO data instead using total active researchers (full-time equivalent) per million people shows a similar trend. Of note, this proxy for active researchers may include non-publishing scientists (private industry, governmental), that are not participating in the strain on scientific publishing in the same way academic scientists are. D) Nevertheless, using UNESCO data the ratio of total articles to total active researchers has gone up substantially since 2019.

**Number of journals by size class, 2002-22**

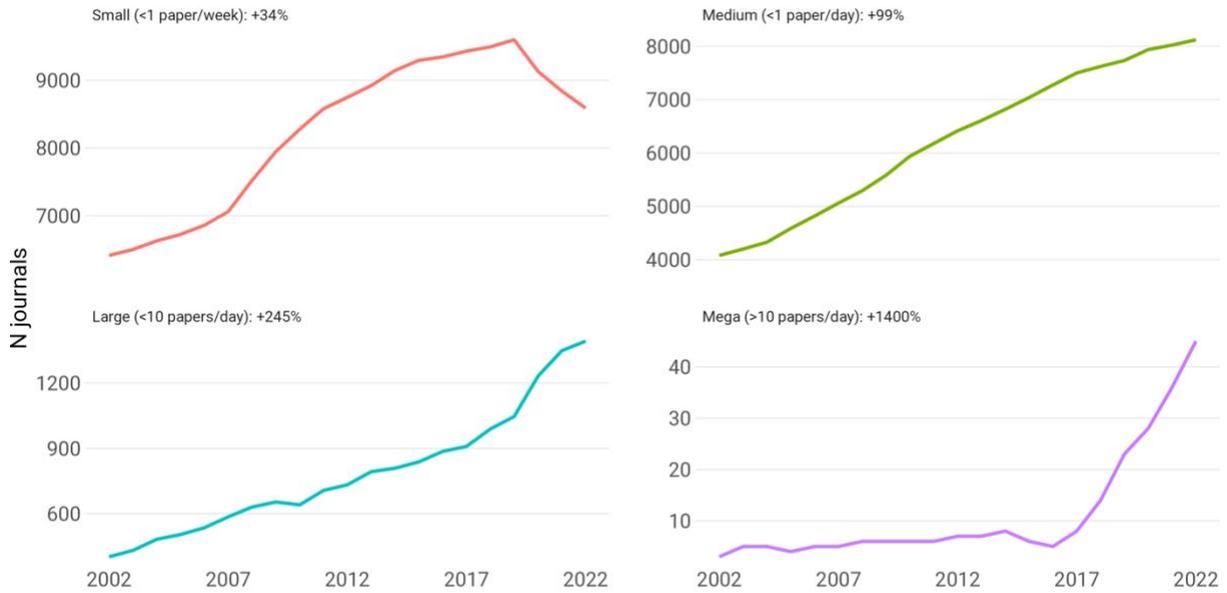

**Figure S3: the rise of megajournals.** We recover trends supporting the article by Ioannidis et al. (2023) who described the "rise of megajournals." Specifically, we see a decline in the number of journals publishing <1 paper/week, but sharp increases in the number of journals publishing over a paper per day. Scientific publishing has therefore been concentrating more and more articles into fewer journals proportionally, which also coincides with a slight decline in the number of journals publishing only a few articles per year.

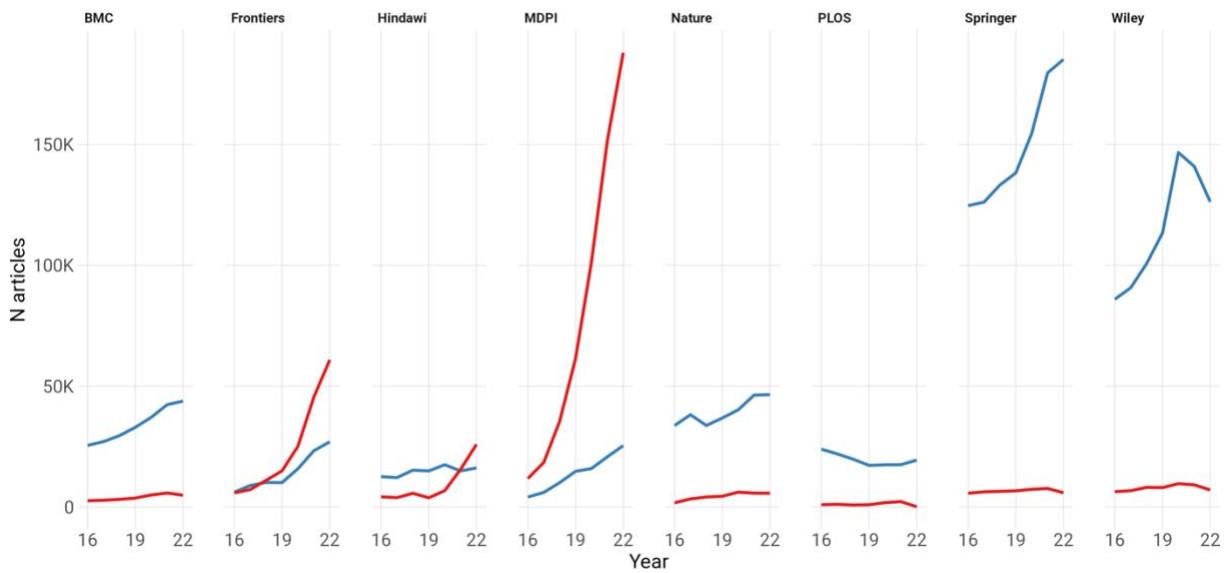

**Number of papers published in regular vs special issues, 2016-22**

Wiley decrease in 2022 likely due to limited coverage of Wiley papers in 2022

Source: data scraped from the publisher's website
Notes: Special issues are called Collections at PLOS and Topics at Frontiers. For MDPI Collections, Sections and Topics not shown.

**Figure S4: proportion of articles published in regular vs. special issues.** Underlying data are the same as in Fig. 2. Line plots are shown to better depict the year-by-year evolving proportion of special issue articles to regular articles. The decline in Wiley articles from 2020-2022 is an artefact of web scraping where total data availability declined in these years. As shown in Fig. 1B, Wiley overall article output increased slightly in recent years.

**Evolution of the share of papers appearing in Special Issues, 2016 to 2022**

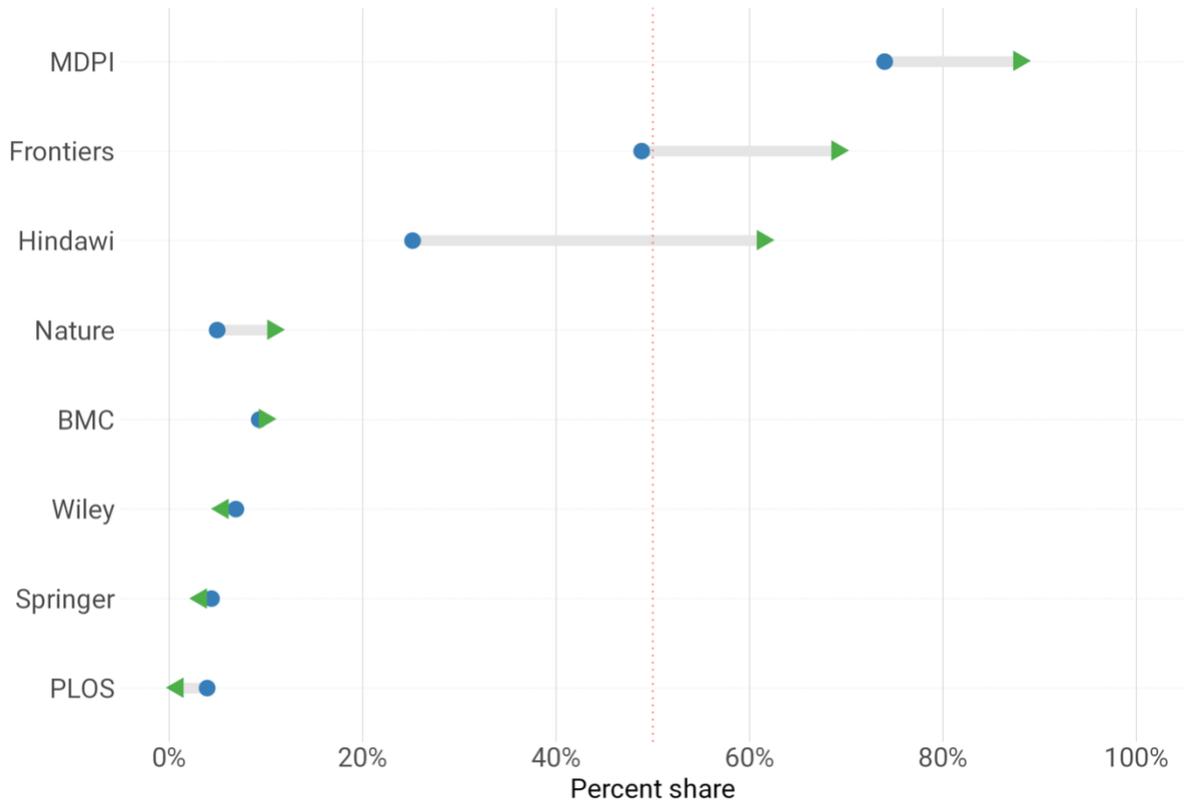

Source: data scraped from the publishers' website
Notes: Special issues are called Collections at PLOS and Topics at Frontiers. For MDPI Collections, Sections and Topics not shown.

**Figure S5: change in special issue between 2016 and 2022.** Certain groups publish the majority of their articles through special issues. Mean proportions of articles published through regular or special issues shown.

**Article heterogeneity in turnaround times by publisher, 2016-22**

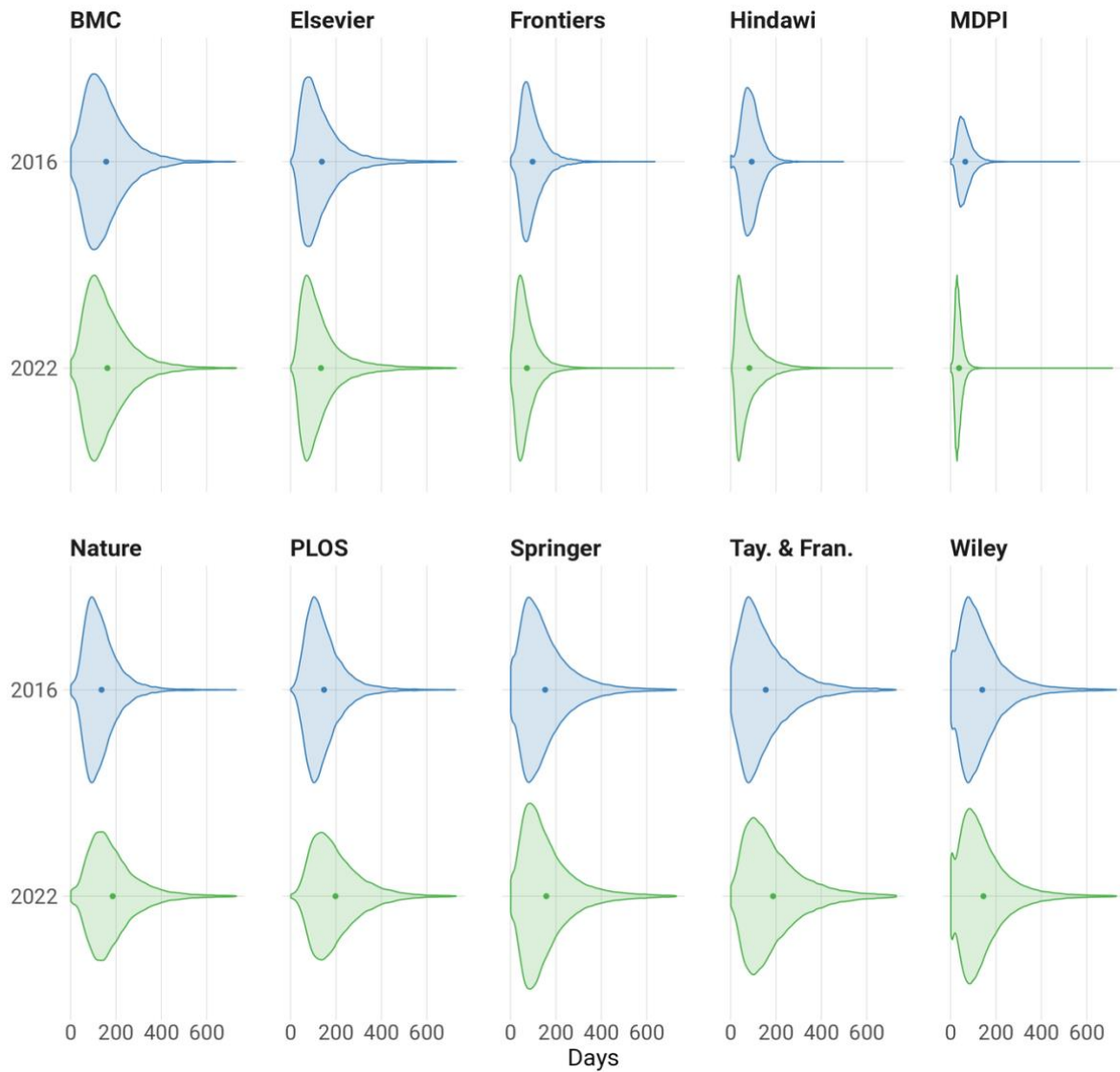

**Figure S6: heterogeneity in journal mean turnaround times by publisher.** Underlying data same as Fig. 3B. Here Violin plots provide an alternate depiction of the density of turnaround time distributions of all articles within their publishing house. "Tay. & Fran." = Taylor & Francis.

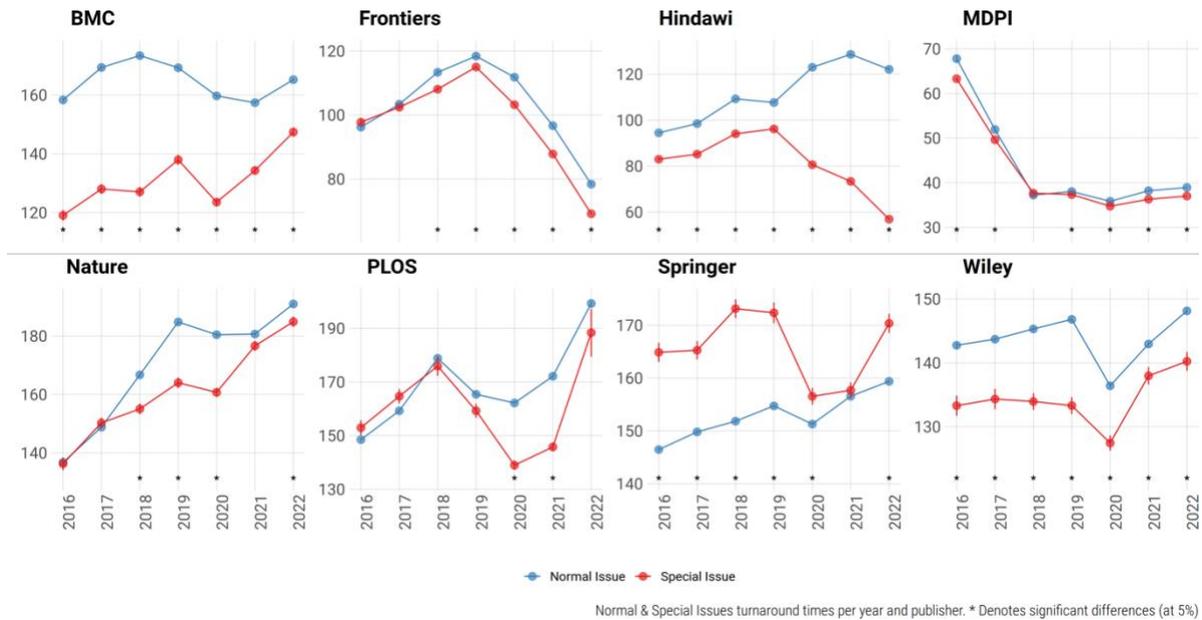

**Figure S7: article turnaround times split by normal or special issue status.** Across publishers, special issue articles have lower turnaround times, often by significant margins (for each year: t-tests corrected with Holm's-Sidak multiple test correction, p < .05 = * along x-axis). The only exception to this trend is Springer, which had higher turnaround times for special issue articles. Of note, the way that special issues are organised can vary across journals and publishers, which could explain the differences in the extent of these trends by publisher. Error bars represent standard error.

**Table S1: descriptive statistics for article turnaround times by publisher in 2016 and 2022.** Q25 and Q75 give interquartile range. SD = Standard Deviation. All values rounded to nearest integer. Of note: the mode of Wiley data is turnaround time = 1, which suggests this dataset (specifically Wiley, which is collected separately from other publishers) may not be fully representative. This parallels a strangeness in Fig. S4 where our scraped data fail to recover the growth of total Wiley articles seen in Fig. 1B. We chose to keep our filter of "turnaround time > 0 days" as the minimum, as in analyses of hundreds of thousands of articles we do expect some genuine outliers, and any cutoff other than turnaround time > 0 days would be defined arbitrarily. Our Wiley data nevertheless suggest that Wiley turnaround time mean and median are over 100 days, which we believe is informative to broad central tendency expectations of Wiley article turnaround times even if mode = 1.

| | 2016 | | | | | | 2022 | | | | | |
|---|---|---|---|---|---|---|---|---|---|---|---|---|
| | ARTICLES | MEAN (SD) | MODE | Q25 | MEDIAN | Q75 | ARTICLES | MEAN (SD) | MODE | Q25 | MEDIAN | Q75 |
| **MDPI** | 20k | 65 (36) | 42 | 41 | 58 | 81 | 271k | 37 (20) | 28 | 25 | 33 | 45 |
| **Frontiers** | 12k | 97 (58) | 63 | 59 | 84 | 120 | 88k | 72 (55) | 49 | 37 | 59 | 91 |
| **Hindawi** | 17k | 93 (43) | 69 | 63 | 88 | 116 | 42k | 83 (67) | 36 | 38 | 61 | 107 |
| **Elsevier** | 58k | 138 (100) | 60 | 71 | 111 | 173 | 72k | 134 (101) | 77 | 68 | 105 | 165 |
| **Wiley** | 91k | 140 (107) | 1 | 68 | 114 | 182 | 131k | 145 (111) | 1 | 70 | 119 | 189 |
| **Springer** | 122k | 152 (113) | 77 | 74 | 124 | 200 | 187k | 157 (116) | 77 | 77 | 127 | 203 |
| **BMC** | 28k | 157 (99) | 112 | 87 | 137 | 204 | 49k | 162 (102) | 112 | 91 | 139 | 209 |
| **Taylor & Francis** | 29k | 155 (117) | 84 | 72 | 124 | 207 | 85k | 187 (130) | 98 | 93 | 154 | 247 |
| **Nature** | 35k | 136 (79) | 105 | 84 | 120 | 170 | 52k | 185 (111) | 133 | 110 | 161 | 234 |
| **PLOS** | 25k | 148 (85) | 97 | 91 | 128 | 182 | 20k | 198 (110) | 147 | 119 | 174 | 252 |

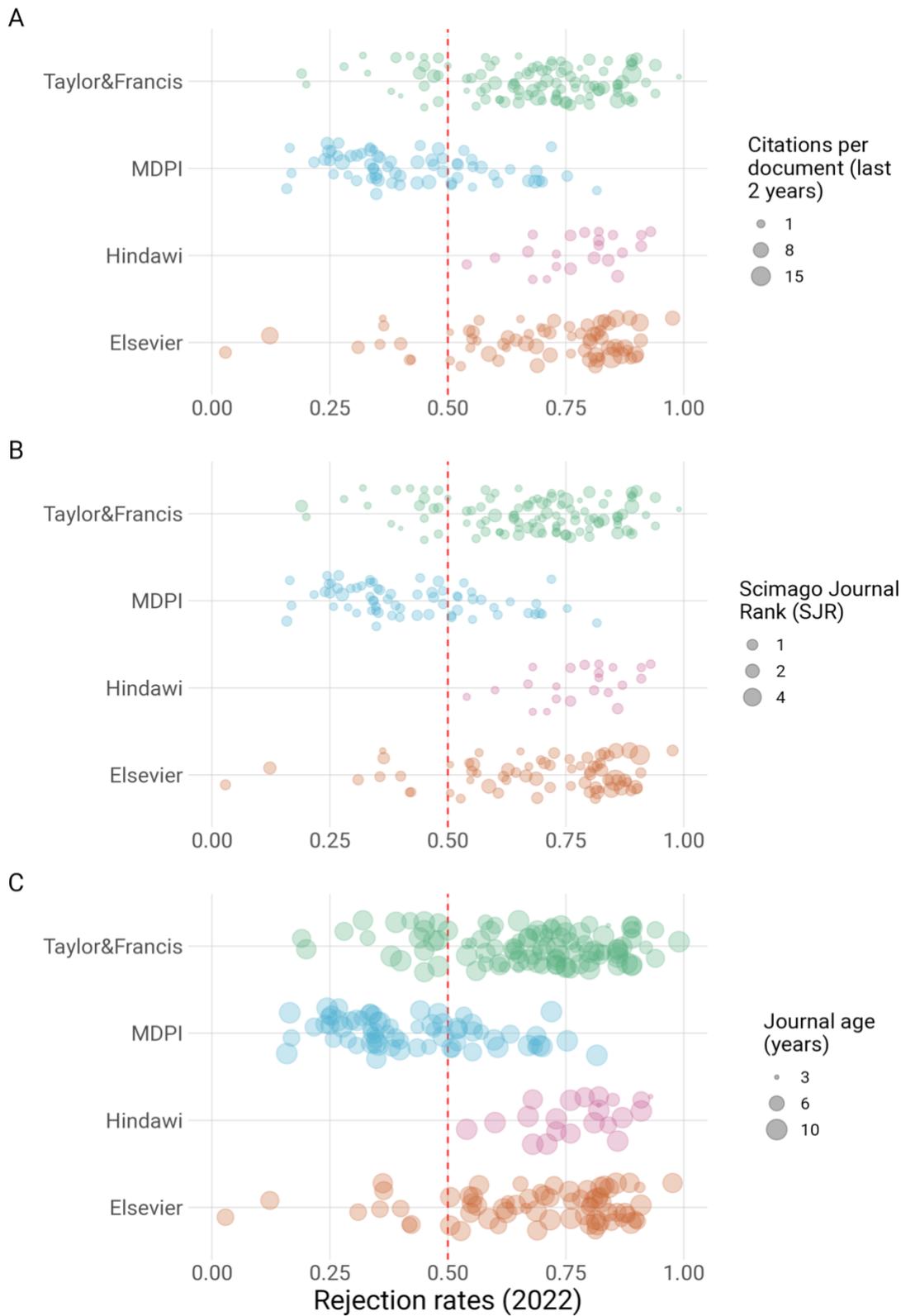

**Figure S8: 2022 rejection rates by publisher, split across different parameters**. Using a general linear model, we found no significant effect of total documents (Fig. 4B), citations per document (A), Scimago Journal Rank (B), or journal age (C) on a journal's 2022 rejection rates across publishers. We chose to investigate young journals (≤ 10 years) to avoid comparing long-established journals to new journals that might have different needs for growth.

## Evolution of rejection rates by relative size of the journal at MDPI, 2016-22

Only journals existing in 2016

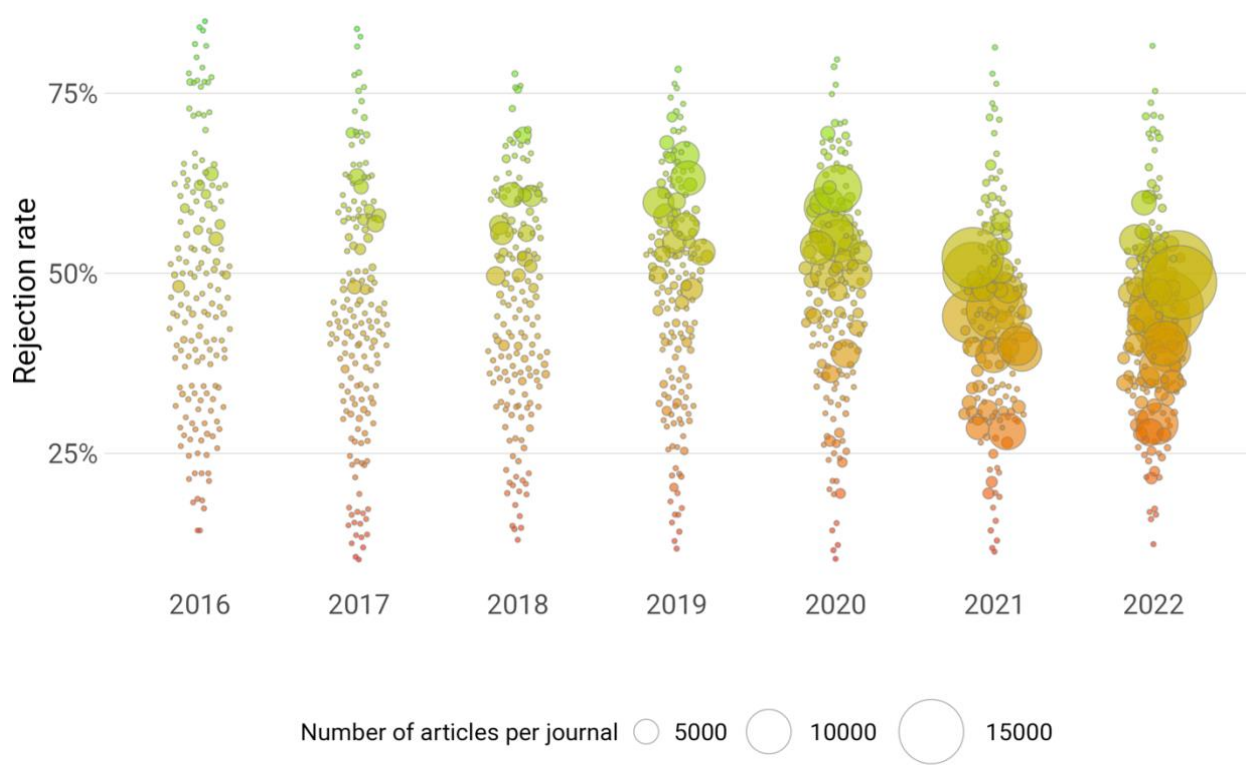

Source: data scraped on the publisher's website

**Figure S9: the decline in MDPI rejection rates is present across journals of different size classes.** A steady decline in rejection rates began between 2019-2020 (Fig. 4A) alongside growth in journal size (larger bubbles here).

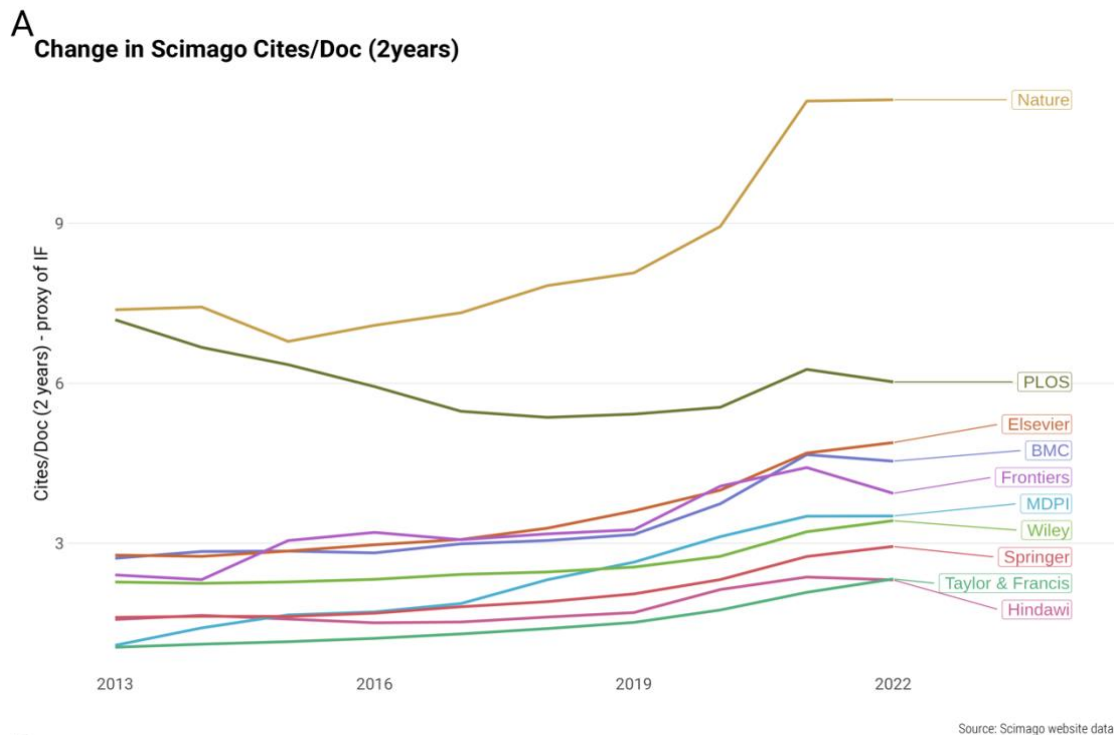

A
**Change in Scimago Cites/Doc (2years)**

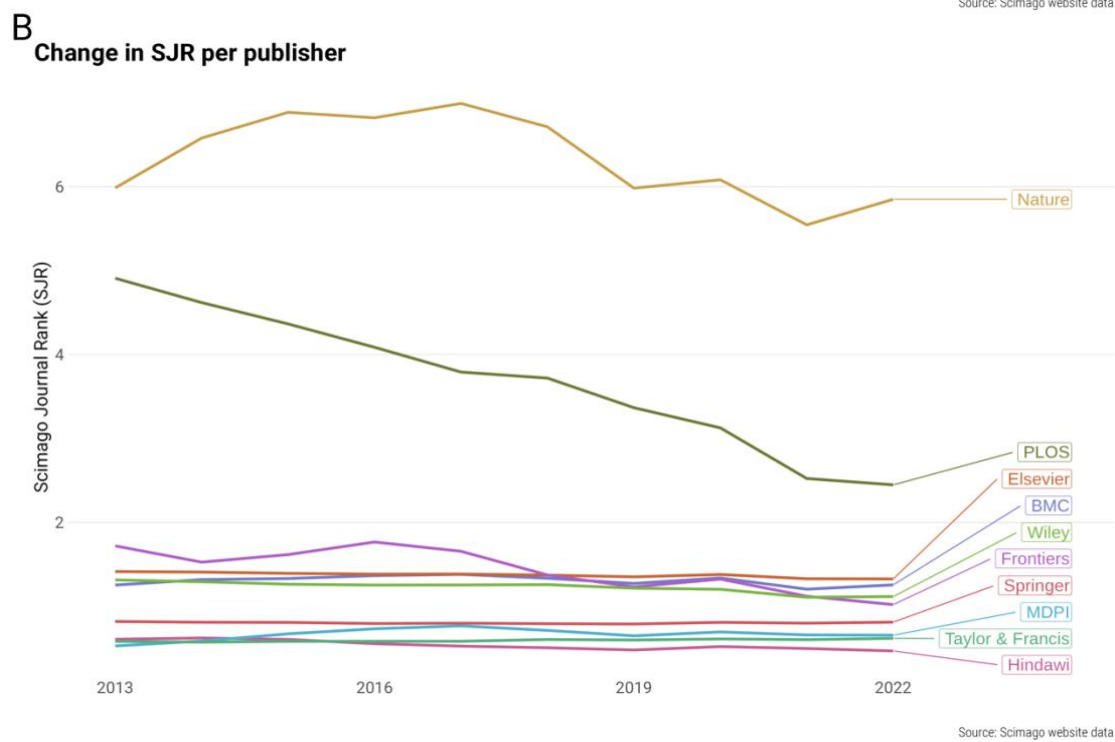

B
**Change in SJR per publisher**

**Figure S10: raw Cites/Doc and SJR informing the Impact Inflation metric.** A) Cites/Doc has been increasing year-over-year across publishers, with a notable uptick beginning after 2019. Here we describe the recent inflation of journal IF (with Cites/Doc as our proxy), suggesting the relative value of a given absolute IF number (e.g. "IF = 3") has decreased more rapidly than in years prior to 2019. B) The SJR has remained relatively constant in recent years, as expected since this metric is normalised for journal size and rate of citable documents generated, rather than raw total documents (Guerrero-Bote & Moya-Anegón, 2012). This suggests that the year-over-year increase in Impact Inflation (Fig. S11) we've observed is due to increased total citations by increasing total articles, but those citations are not weighted as "prestigious" in a network-adjusted metric compared to pre-2019 years.

## Impact inflation by journal size 1999-2022

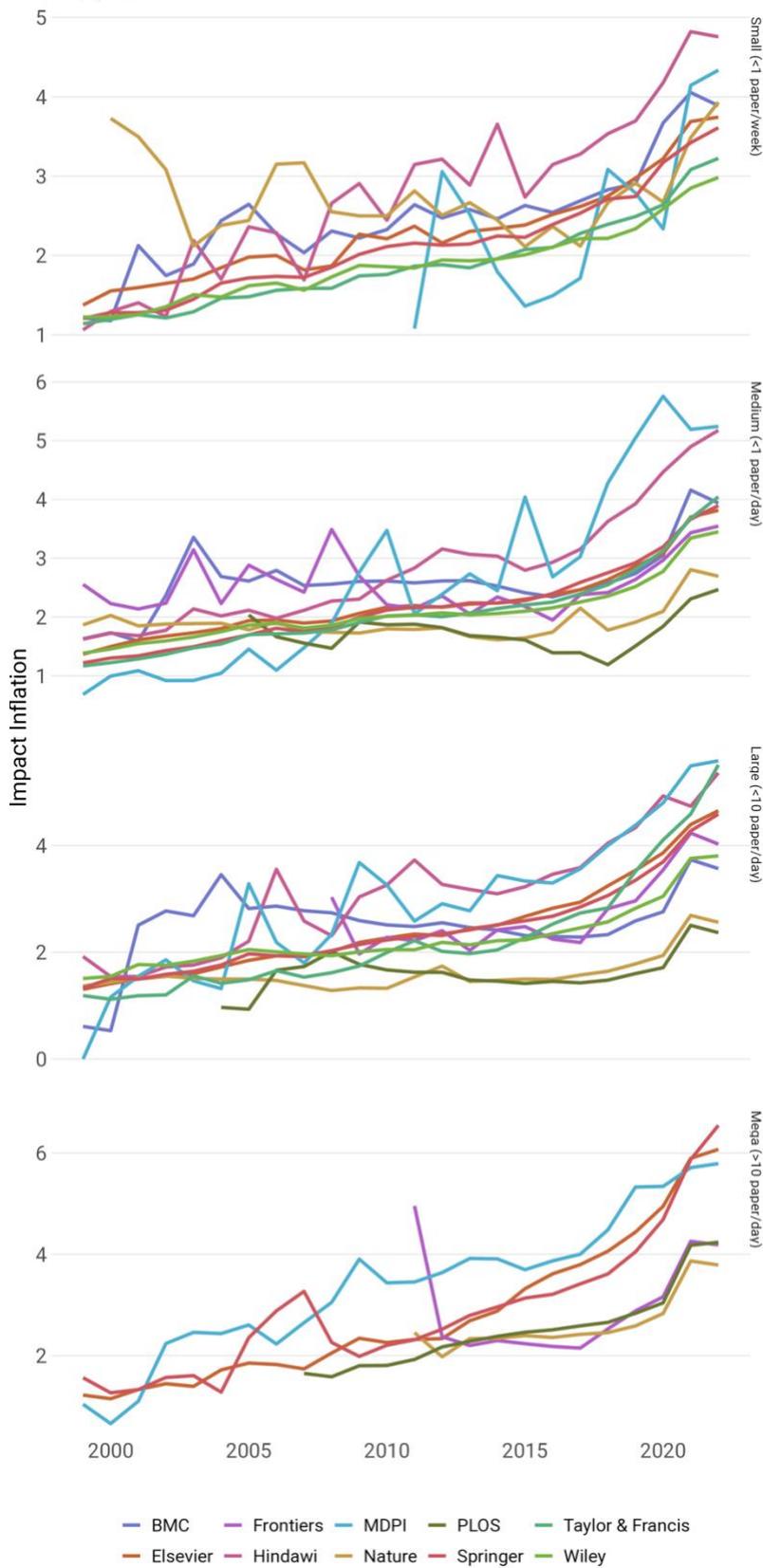

**Figure S11: there has been a universal increase in Impact Inflation independent of journal size across all publishers.** Also see Fig. S14A.

**References per document, 2016-22**

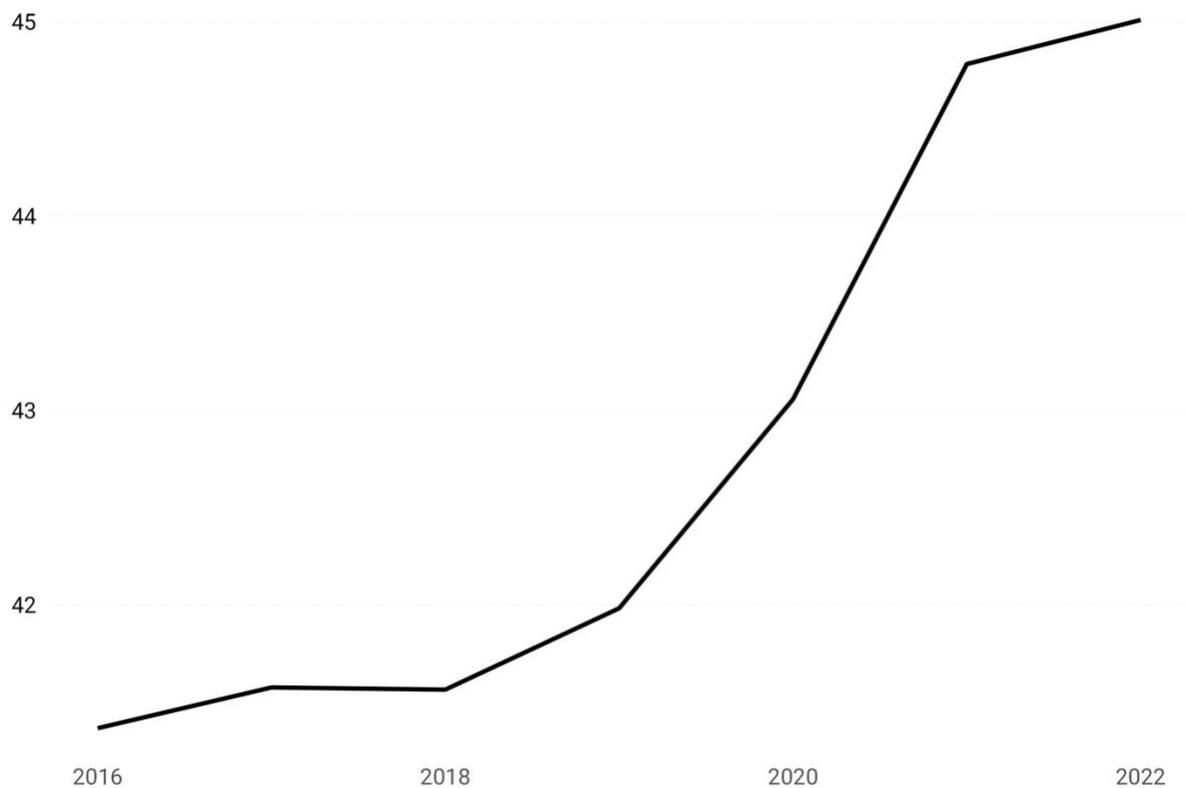

**Figure S12: a partial contributor to the increasing total citations being generated is an exponential increase in references per document between 2018-2021.** As such, not only is more work being produced (total article growth), but that work is also proportionally generating more citations than articles would be in past years. Here we will note that this growth overlapped the COVID-19 pandemic, which provided an excess of potential writing time to scientists. However, growth in references per doc began already between 2018 and 2019, suggesting the effect of COVID-19 cannot fully explain this change. The ensuing year of 2020 also coincides with acceleration of articles published through special issues (Fig. S4). Thus the growth in references per document is correlated both with a burst of special issue publishing, and world events. References per document also continued to increase between 2021 and 2022 despite measures around COVID-19 relaxing in 2022 – albeit there is indeed a marked decrease in the rate of growth. A full understanding of the influence of COVID-19 on this growth in references per document, and how much references per document explains the universal increase in impact factor (Fig. S10, S11) will await data from 2023 where the impact of COVID-19 is further lessened, and normalised for the significant delistings that Clarivate performed in March 2023 that have had a marked effect on the calculation of impact factor (Fig. S13A).

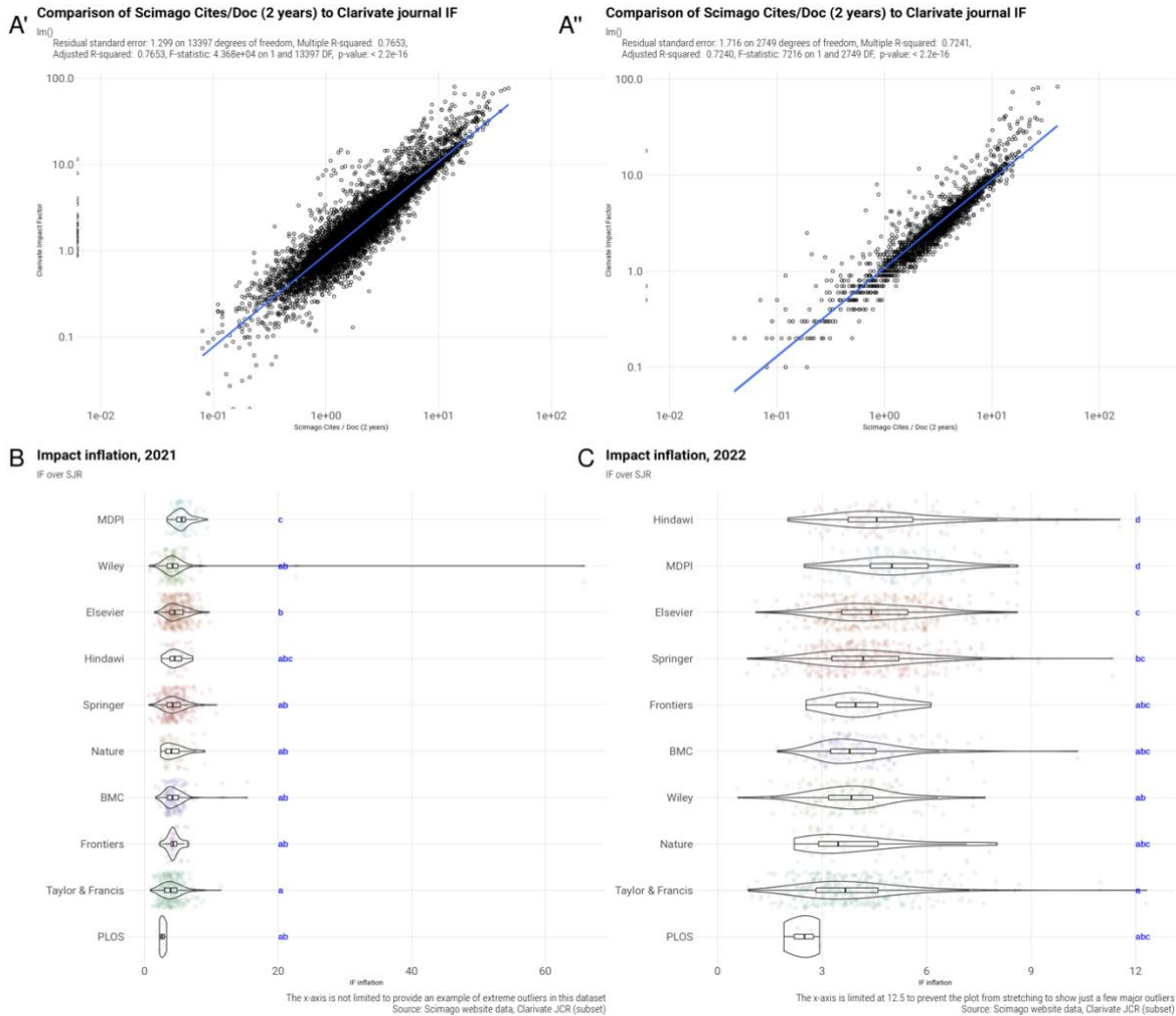

**Figure S13: validation of Scimago Cites/Doc (2 years) as a proxy of Clarivate journal IF.** A) Prior to 2022, "Cites/Doc (2 years)" and Clarivate IF have a correlation of *adj-R²* = 0.77 (A'), but due to mass delistings by Clarivate (but not Scimago) affecting 2022 journal IFs, there was a decoupling of this correlation for 2022 (A'' : *adj-R²* = 0.72). Regardless, Cites/Doc (2 years) informed by the Scopus database is a good proxy of Clarivate Web of Science IF. B-C) Impact Inflation calculated using a subset of Clarivate IFs we could download for our publishers of interest in 2021 (B) and 2022 (C). In both years, MDPI has significantly higher Impact Inflation compared to all other publishers except Hindawi. Here we leave an example in (B) of what is meant by "major outliers" in Fig. 5, to show that plotting the full x-axis range does not change trends, but is aesthetically disguising.

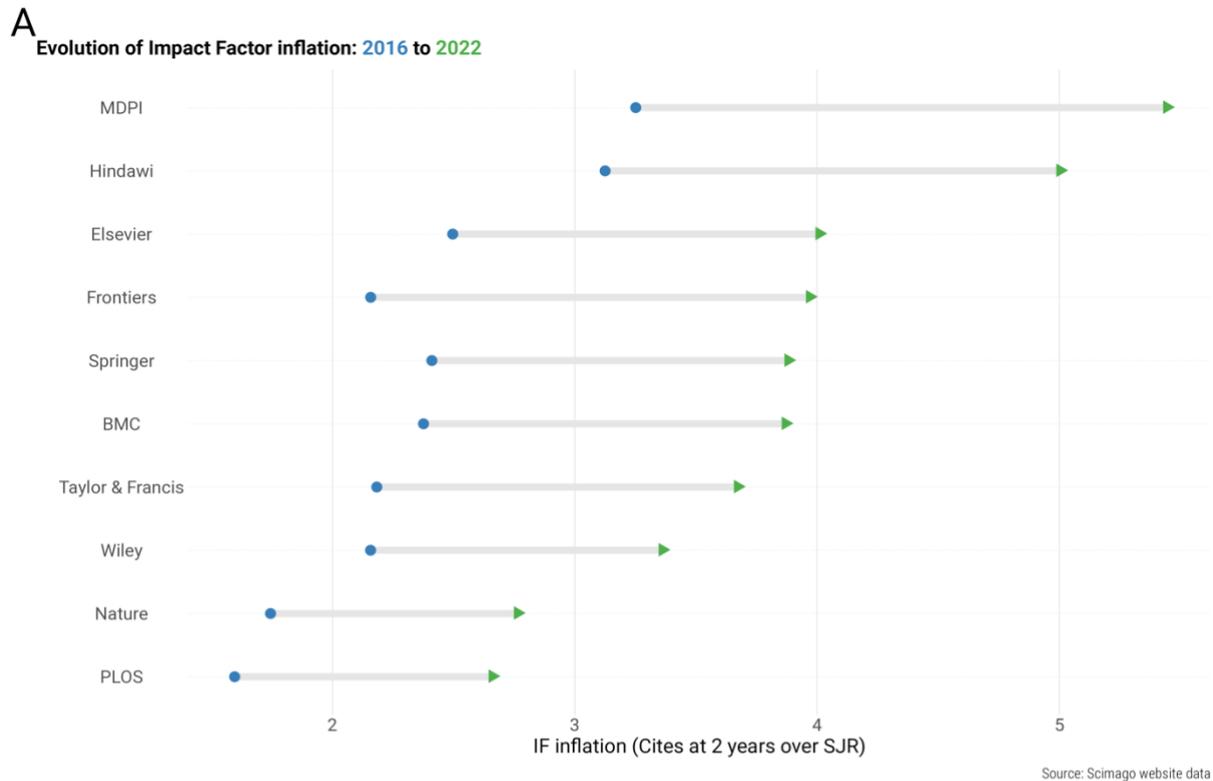

A

**Evolution of Impact Factor inflation: 2016 to 2022**

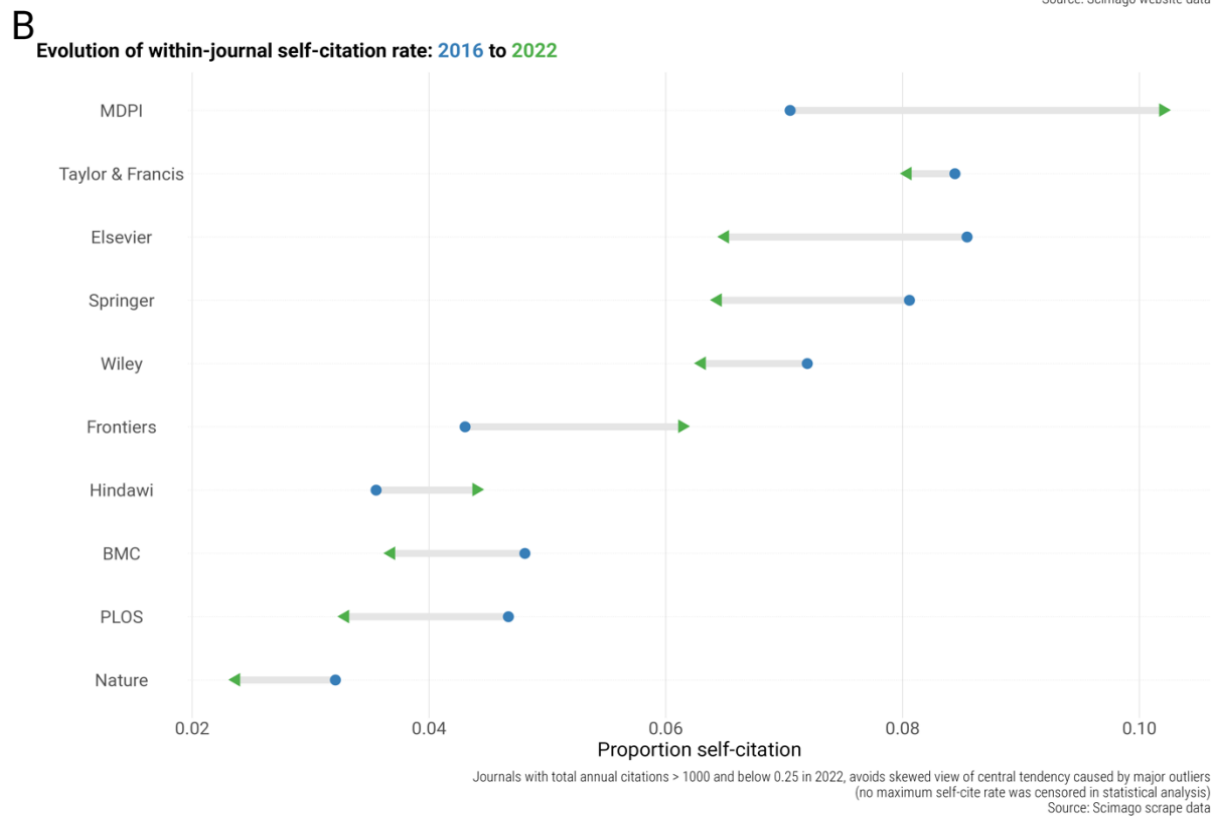

B

**Evolution of within-journal self-citation rate: 2016 to 2022**

**Figure S14: evolution of Impact Inflation and within-journal self-citation between 2016 and 2022.** A) Impact Inflation has increased universally across publishers (absolute values summarised in Table 2). B) Within-journal self-citation has increased in recent years specifically for publishers that grew through use of the special issue model of publishing: MDPI, Frontiers, and Hindawi. Notably, MDPI has higher self-citation rates than any other publisher, exceeding previous highs from 2016 (Elsevier, Taylor & Francis) by over one percentage point.

## Within-journal self citation rate, 2021

Does not include journals citing across each other

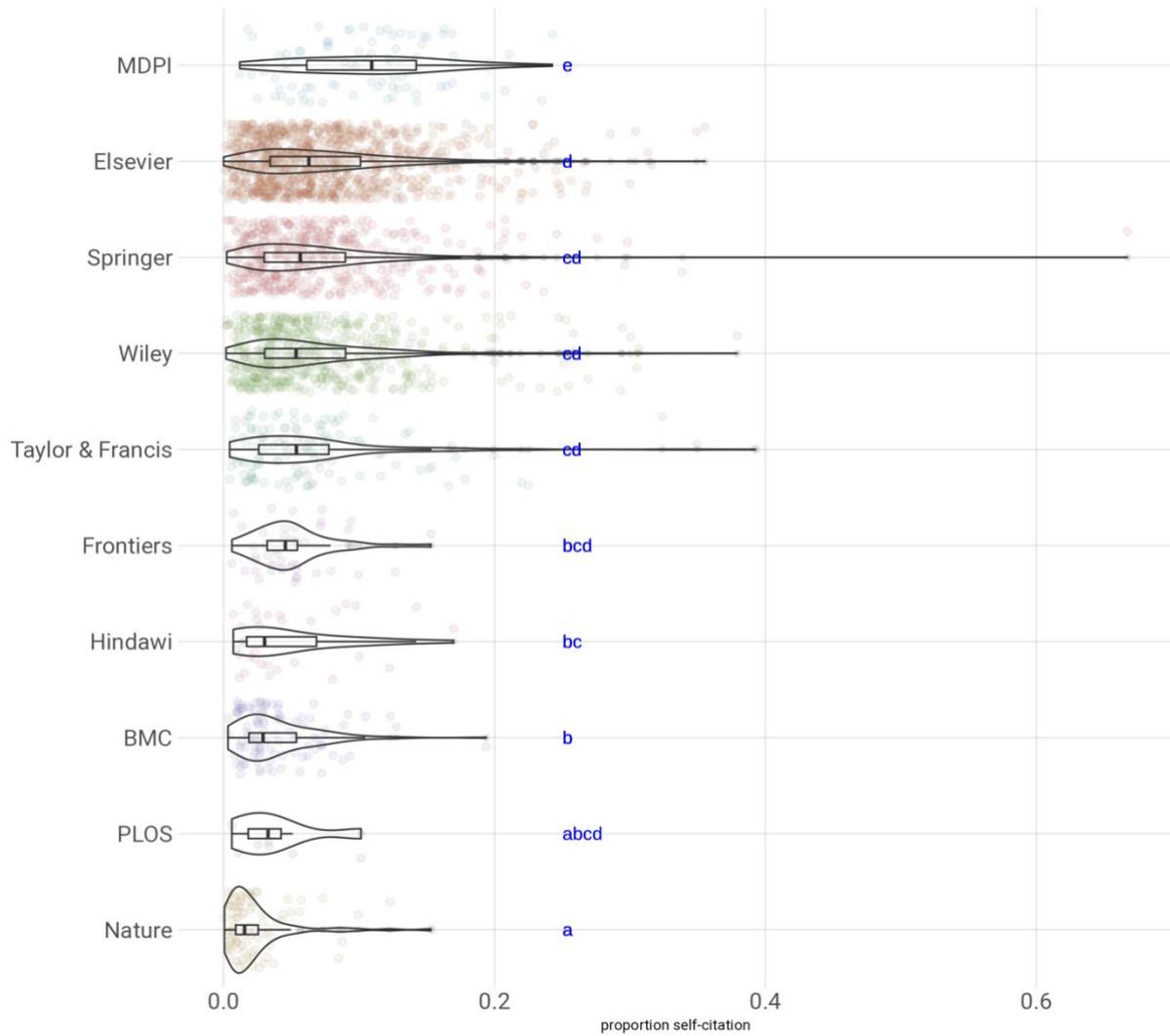

shows journals with total annual citations > 1000, and the x axis is cut off at 0.25 to prevent the plot from stretching due to a few major outliers
Source: Scimago scrape data

**Figure S15: within-journal self-citation rates from 2021, supporting the trend in 2022 that MDPI uniquely has significantly higher self-citation rates compared to all other publishers.** A difference between 2021 and 2022 is that in 2022, MDPI and Taylor & Francis were not significantly different (*P* > .05). In 2021, this difference was significant (P = 3e-7).

**MDPI**

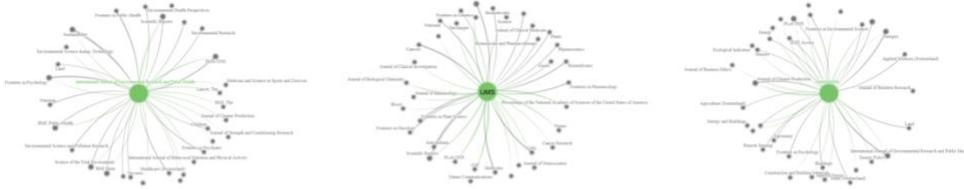

**Hindawi**

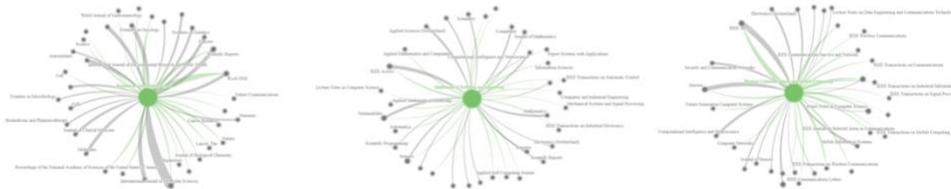

**Frontiers**

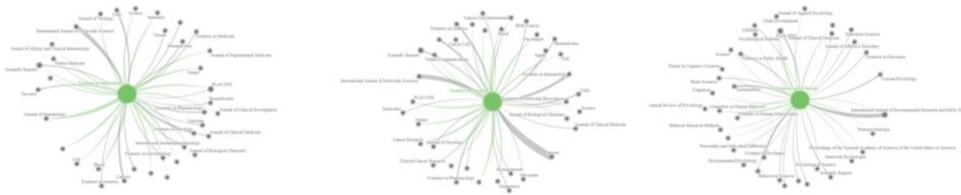

**BMC**

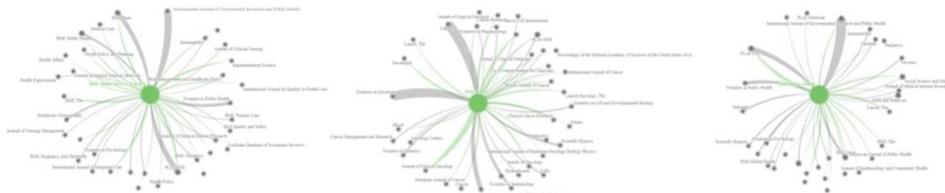

**PLOS**

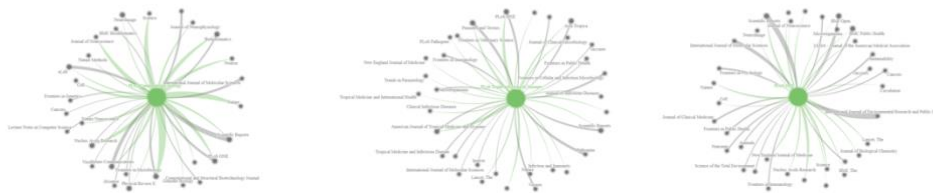

**Figure S16: example citation networks of single journals from Scimago.** Journals were selected from the largest journals by publisher. Journal citation reciprocity depicted with grey arrows for incoming citations, and green arrows for outgoing citations. MDPI journals make up large fractions of the total incoming citations of their own journals, uniquely true of MDPI and not other publishers in our analysis. This result is in keeping with MDPI themselves, who reported a ~29% within-MDPI citation rate (Fig. S18). High rates of Impact Inflation of Hindawi journals may come from disproportionate citations received from MDPI journals. For instance, a plurality of citations to BioMed Research International (row 2, column 1) come as large chunks (thick grey arrows) from MDPI journals (International Journal of Molecular Sciences, International Journal of Environmental Research and Public Health, Nutrients, Antioxidants, Cancers, etc…). A similar pattern is seen for Mathematical Problems in Engineering (row 2, column 2): Sustainability, Mathematics, Applied Sciences (Switzerland), Symmetry, Sensors, etc… Because the Scimago Journal Rank metric has an upper limit on the prestige a single source can provide, the large number of citations individual MDPI journals are exporting may be an important factor leading to universal trends in Impact Inflation. *A full-resolution version of this figure is available online at:* [https://figshare.com/articles/figure/The_strain_on_scientific_publishing_figures_/24203790](https://figshare.com/articles/figure/The_strain_on_scientific_publishing_figures_/24203790).

**Table S2: Change in submitted papers relative to the previous month for the 25 largest MDPI journals.** On March 23[rd] 2023 Clarivate announced the delisting of the MDPI flagship journal International Journal of Environmental Research and Public Health (IJERPH), as well as Journal of Risk and Financial Management (JRFM). Following this, submissions to IJERPH plummeted by 73 percentage points in April 2023 compared to March 2023, which already showed a slowdown overall compared to February 2023. Moreover, submissions to MDPI journals in general were down in April 2023 across the board compared to March. A similar pattern was seen in early 2022 following the Chinese Academy of Science release of their "Early Warning Journal List" trial published Dec 31[st] 2021, which featured multiple MDPI journals. These patterns demonstrate that external authorities, such as Clarivate or national academies of science, can have profound impacts on author submission behaviour, despite opaque methodologies surrounding their decisions to list or delist journals.

**Monthly % change in submitted papers – 25 largest MDPI journals**
Over all 98 journals with an Impact Factor as of February 2023 and by journal

| JOURNAL | N | 2022 | | | | | | | | | | | | 2023 | | | | | | | | | | | |
|---|---|---|---|---|---|---|---|---|---|---|---|---|---|---|---|---|---|---|---|---|---|---|---|---|---|
| | | JAN | FEB | MAR | APR | MAY | JUN | JUL | AUG | SEP | OCT | NOV | DEC | JAN | FEB | MAR | APR | MAY | JUN | JUL | AUG | SEP | OCT | NOV | DEC |
| Overall | 287573 | NA | -9.77 | 12.85 | -3.45 | 0.53 | 0.19 | -1.88 | 7.36 | 8.67 | 10.82 | 5.43 | 2.15 | 0.31 | 4.29 | 4.90 | -21.13 | -3.96 | -0.29 | -0.61 | -1.57 | 0.20 | -1.47 | 3.11 | -7.75 |
| ijerph | 17278 | NA | -8.77 | 18.78 | 2.57 | -2.75 | 2.66 | -2.11 | 4.97 | 12.22 | 10.58 | 7.53 | 1.06 | 1.23 | 5.72 | -17.76 | -73.12 | -3.09 | -23.21 | -8.97 | -2.55 | -16.67 | -8.09 | -0.49 | -10.57 |
| sustainability | 17242 | NA | -6.03 | 18.18 | -4.19 | 2.84 | -7.71 | -1.25 | 17.01 | 1.99 | 15.28 | 5.20 | -1.61 | -3.04 | 11.13 | 23.61 | -19.57 | -2.33 | 9.50 | -5.56 | -1.62 | 5.96 | 1.71 | -9.49 | -17.51 |
| ijms | 16140 | NA | -15.63 | 12.52 | 0.45 | 0.33 | 5.62 | -6.21 | 6.01 | 17.66 | 12.10 | 6.27 | 2.30 | -1.92 | 6.59 | -0.54 | -18.61 | -1.38 | 2.59 | -2.69 | 2.99 | -3.50 | -5.10 | 9.92 | -0.13 |
| applsci | 12890 | NA | -14.78 | 4.35 | -13.43 | 10.81 | -0.70 | -1.35 | -7.50 | 4.67 | 13.74 | 8.55 | -2.43 | 10.51 | 4.20 | 11.82 | -11.55 | -2.22 | -4.51 | -0.53 | -4.13 | 3.93 | 1.38 | 7.99 | -5.25 |
| sensors | 10084 | NA | -7.21 | 14.64 | -8.49 | -1.33 | 1.18 | -2.11 | 3.19 | 13.06 | 7.72 | 6.47 | 13.83 | -8.53 | 2.69 | 5.57 | -12.83 | -14.23 | -1.69 | -0.65 | -4.12 | 0.60 | -0.45 | 11.37 | -17.04 |
| energies | 9588 | NA | -11.50 | 10.02 | -0.98 | 6.71 | -4.89 | -3.18 | 4.05 | 1.54 | 30.41 | 4.04 | -0.76 | -5.75 | 5.53 | -9.77 | -5.94 | -11.94 | -2.63 | 4.42 | -1.57 | -2.39 | -5.39 | -6.73 | 1.94 |
| molecules | 9066 | NA | -7.71 | 11.86 | -8.69 | 4.00 | -3.93 | -0.19 | 3.72 | 18.12 | 10.67 | 5.84 | 8.11 | 2.89 | -6.99 | 2.63 | -14.36 | -15.79 | 0.87 | 1.81 | 13.69 | -0.29 | 6.78 | -0.55 | -1.85 |
| materials | 9046 | NA | -15.02 | 14.82 | -11.28 | -0.54 | 2.49 | -0.11 | 10.16 | 9.70 | 9.63 | 2.08 | 5.09 | -2.38 | -3.89 | 0.24 | -7.36 | -9.06 | -3.67 | 1.17 | -5.21 | -1.73 | -5.59 | 3.73 | -9.51 |
| jcm | 7299 | NA | 0.31 | 7.22 | -3.98 | 4.94 | 10.07 | -7.09 | 5.52 | 2.53 | 0.68 | 2.36 | 10.56 | 23.81 | 11.27 | 18.53 | -26.60 | -11.39 | 16.23 | -4.53 | 9.94 | -1.79 | -12.43 | 16.88 | -1.78 |
| remotesensing | 6416 | NA | -5.46 | 11.14 | -0.27 | 3.29 | -0.80 | 0.18 | 14.87 | 1.16 | 2.53 | 5.16 | -5.61 | -2.18 | 1.08 | 9.67 | -15.49 | 1.64 | -4.69 | 3.73 | 0.82 | 2.43 | -0.79 | -0.08 | -6.07 |
| cancers | 6252 | NA | -8.09 | 7.20 | -5.31 | 0.42 | -2.21 | 6.90 | 12.50 | 9.05 | 7.81 | 1.07 | 11.92 | -6.81 | 3.83 | -5.01 | -22.87 | -2.57 | 1.85 | 3.93 | 1.11 | -1.91 | -4.93 | 8.80 | -5.48 |
| polymers | 5573 | NA | -4.59 | 8.70 | -7.84 | -6.10 | 12.01 | 5.62 | 11.81 | 3.42 | 19.71 | 2.64 | -5.62 | -1.86 | -1.77 | 13.38 | -23.50 | -2.52 | -6.09 | -0.65 | -4.08 | 5.10 | -0.65 | -2.93 | -10.57 |
| nutrients | 5276 | NA | -0.14 | 5.90 | 3.50 | 1.00 | -4.58 | 0.91 | 8.37 | 3.92 | 12.11 | 5.61 | 0.19 | 1.83 | 1.23 | 5.70 | -19.19 | 6.78 | -6.05 | 13.41 | -1.73 | -0.59 | 6.20 | 4.73 | -9.03 |
| mathematics | 4884 | NA | -13.35 | 17.10 | 15.50 | -4.73 | -6.08 | 1.62 | 9.87 | 10.43 | 14.35 | -13.54 | 5.58 | 0.92 | -3.74 | 14.67 | -13.32 | 1.56 | -2.74 | -8.17 | -4.69 | 13.76 | -8.83 | 0.77 | -19.40 |
| nanomaterials | 4434 | NA | -14.68 | 16.28 | 11.83 | -3.88 | 4.95 | 0.35 | -2.26 | 9.09 | 8.17 | 3.32 | 5.16 | -13.27 | -4.70 | -5.66 | -8.51 | -5.50 | -17.00 | -10.78 | -0.73 | -0.25 | -5.65 | -10.16 | 17.39 |
| electronics | 4272 | NA | -8.44 | 21.84 | -7.73 | -9.45 | -0.98 | 5.96 | 10.32 | 10.20 | 7.41 | 19.68 | 8.16 | 8.32 | 10.45 | 19.02 | -20.73 | -5.11 | -3.42 | 3.43 | -4.25 | 3.47 | 4.29 | 8.53 | -15.82 |
| water | 4147 | NA | -11.16 | 1.80 | -3.53 | -0.55 | 4.41 | -16.55 | 19.83 | 2.29 | 12.05 | 2.00 | -0.60 | -3.18 | 4.38 | 3.00 | 5.53 | -12.00 | -1.72 | 1.91 | 0.00 | -6.26 | 9.68 | 0.00 | -6.39 |
| foods | 4107 | NA | -14.14 | 16.27 | -10.92 | 3.64 | 7.02 | 12.95 | 4.13 | 13.95 | 7.09 | 1.81 | -8.04 | -1.16 | 11.18 | 4.33 | -15.59 | -1.24 | -0.54 | -6.33 | 9.46 | 6.00 | 3.50 | -0.68 | -4.36 |
| cells | 4076 | NA | -8.66 | 14.22 | -2.45 | -10.06 | -5.81 | -1.60 | 23.90 | 9.55 | -3.93 | 18.33 | 5.71 | 2.84 | 7.88 | -6.15 | -34.15 | 9.75 | -2.46 | -0.97 | -2.35 | -11.82 | -0.23 | 13.90 | -15.20 |
| animals | 3623 | NA | -0.85 | 23.44 | 19.20 | 1.73 | -0.84 | -6.62 | 15.79 | 15.22 | 3.95 | 7.76 | 1.99 | -0.74 | 13.70 | -5.02 | -8.91 | 7.63 | 3.54 | -0.89 | -0.75 | -10.14 | 9.08 | -1.21 | |
| plants | 3611 | NA | -16.07 | 22.77 | -11.44 | 2.74 | 4.95 | -7.44 | 7.06 | 4.58 | 19.79 | 5.41 | 7.91 | 5.27 | -5.01 | 7.97 | -18.69 | 0.44 | -5.39 | 9.09 | -10.03 | 5.49 | -1.93 | 6.22 | -9.86 |
| biomedicines | 3249 | NA | -2.23 | 11.59 | -2.04 | 1.14 | 5.62 | -25.00 | 12.29 | 6.11 | 3.77 | 9.37 | -7.17 | 0.00 | 17.70 | 16.16 | -16.67 | 1.82 | 2.92 | 11.99 | -0.85 | -14.49 | 9.80 | 10.74 | -12.16 |
| agronomy | 3233 | NA | -17.79 | 7.56 | -2.93 | -1.01 | -5.08 | 5.35 | -7.11 | 27.57 | -1.20 | 3.12 | -0.51 | -0.34 | -1.87 | 15.40 | -15.59 | -1.24 | -0.54 | -6.33 | 9.46 | 6.00 | 0.33 | -5.47 | 0.53 |
| diagnostics | 3219 | NA | -11.89 | 14.46 | 5.26 | -16.20 | 10.26 | -5.19 | 7.76 | 3.60 | 6.13 | 20.62 | 7.19 | 10.88 | 0.13 | 10.60 | -17.48 | 5.88 | -7.36 | -3.75 | -7.63 | 9.95 | -9.51 | -2.54 | -5.74 |
| pharmaceutics | 2858 | NA | -10.51 | 12.01 | 1.17 | -3.00 | -8.31 | 9.59 | 7.33 | 8.15 | 14.26 | 14.80 | -1.09 | 3.45 | 0.76 | 7.53 | -19.33 | -10.24 | -5.42 | -6.13 | 0.00 | -11.33 | 2.95 | 0.95 | -8.51 |

Source: data scraped from the publisher's website by @paolocrosetto

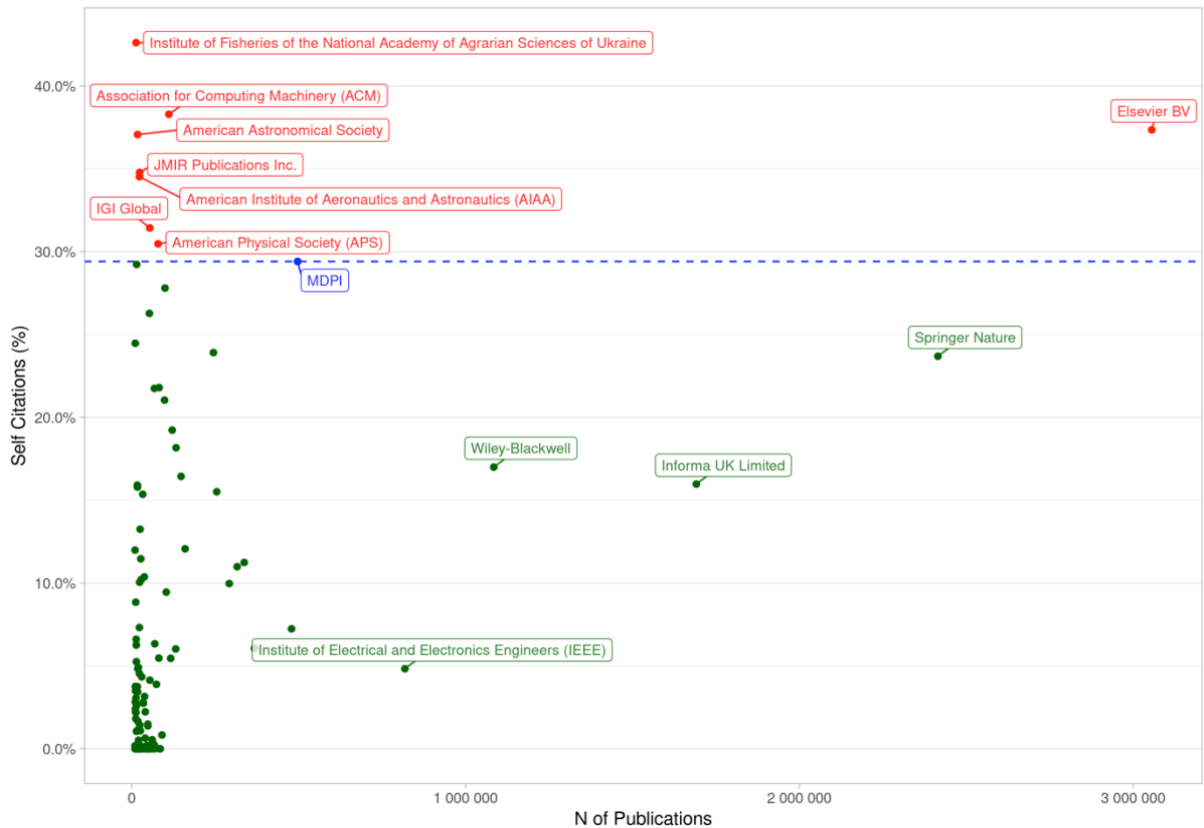

**Figure S17:** analysis of within-publisher self-citation rate performed by MDPI (MDPI, 2021) in response to Oviedo-Garcia (Oviedo-García, 2021). The original interpretation of this figure, as presented by MDPI, is: "It can be seen that MDPI is in-line with other publishers, and that its self-citation index is lower than that of many others; on the other hand, its self-citation index is higher than some others." Our data in Fig. 5B (2022) and Fig. S16 (2021) suggest instead that established MDPI journals receiving >1000 citations per year have uniquely high rates of within journal self-citation, which are significantly different from other publishers. This filter for only journals receiving >1000 citations is key, as due to the growth of MDPI journals in recent years, not including this caveat can give the false impression that MDPI, overall, has comparable rates of within-journal self-citation due to the many recent journals with relatively few articles that cannot easily cite themselves (but *can* cite other MDPI journals).

# Supplementary Materials and Methods

## Additional data considerations

*Global researcher statistics*

We ran analyses with or without the category "Arts & Humanities" comprising ~3% of total articles in our filtered Scimago data in any given year, and 8-11% of OECD PhD graduates by year. These inclusion/exclusion criteria do not change the results of analyses relating to Figures 1 and 5, or supplementary figures using the Scimago dataset.

*Grouping of publishers per Scimago labels*

Publisher labels in Scimago were aggregated according to key "brand" names such as "Elsevier" or "Springer"; e.g. Elsevier BV, Elsevier Ltd and similar were aggregated as Elsevier, or Springer GmbH & Co, Springer International Publishing AG as "Springer."

This does not entirely capture the nested publishing structures of certain "publishers" per Scimago labels. At the time of writing, Elsevier and Springer are both publishers who own >2500 journals according to self-descriptions. However, our dataset only assigns ~1600 journals to these publishers in 2022 (Fig. S2). Reasons for this discrepancy between self-reported numbers and our aggregate numbers come from smaller, but independent, publisher groups operating under the infrastructure of these larger publishing houses. Two examples: **1)** Cell Press (> 50 journals) is owned by Elsevier, **2)** both BioMed Central (BMC, >200 journals) and Nature Portfolio (Nature, >100 journals) are owned by Springer. Ultimately, we decided that publishing houses large enough to distinguish themselves with their own licensed names were managed and operated sufficiently independently from their parent corporations to be kept separate. Nevertheless, our dataset aggregates the majority of Elsevier and Springer journals under their namesakes, and so we feel the data we report are a representative sampling, even if we caution that interpretation of trends in "Elsevier", "Nature", or other publishers, should consider this caveat regarding nested publisher ownership.

Our goal of characterising strain meant that we had to focus on publishers that were sufficiently "large" as far as our strain metrics were concerned. We included publishers like Hindawi and Public Library of Science (PLOS) because they were uniquely 'large' in terms of certain business models. PLOS is the largest publisher in terms of articles per journal per year (Fig. 1C), while Hindawi is a major publisher in terms of publishing articles under the Special Issue model (Fig. 2) that is also of current public interest (Quaderi, 2023). We also retained BMC and Nature as independent entities in our study, as these publishers offer relevant comparisons among publishing models. BMC is a for-profit Open Access publisher that operates hundreds of journals, much like Hindawi, Frontiers, and MDPI. Nature is a hybrid model publisher that includes paywalled or Open Access articles, publishes more total articles than BMC (Fig. 1B), and was a distinct publishing group for which we could collect systematic data on Special Issue use and turnaround times (Fig. 2, Fig. 3). We were also able to collect a partial sampling of those data from Springer, but to merge the two would have caused Nature to contribute a strong plurality of trends in Springer data in Fig. 2 and Fig. 3, obscuring the trends of both this nested publishing house and of the remaining majority of Springer journals: indeed the proportion of Special Issues (Fig. 2), turnaround times (Fig. 3), Impact Inflation and self-citation (Fig. 5) of Nature is significantly different from other Springer journals, sometimes by a wide margin.

In some cases, journal size was also a relevant factor for comparisons across publishers. As emphasised by the high number of articles per journal by PLOS, MDPI, and Frontiers (Fig. 1C), some publishers publish hundreds to thousands of articles per journal annually, while others publish far less. The age of journals was also tied to this article output, as newer journals publish fewer articles, but grow to publish thousands of articles annually in later years. We therefore considered journal size throughout (Fig. S3), and in metrics like self-citations, which were censored for only journals receiving at least 1000 citations per year. These filters were applied to ensure comparisons across journals and publishers were being made fairly: for instance, small journals have fewer articles to self-cite to, and highly specific niche journals may have high rates of self-citation for sensible reasons. This was especially important for comparisons at the publisher level, as some publishers have increased their number of journals substantially in recent years (Fig. S2), meaning a large fraction of their journals are relatively young and less characteristic of the publisher's trends according to their better-established journals.

*Rejection rates*

The analysis of rejection rates comes with important caveats: these data come from non-standardised data sources (each publisher decides how rejection rates are reported) and we use voluntarily-reported rather than systematic data (volunteer bias).

In most cases, publishers track the total submissions, rejections, and acceptances over a set period of time. This can sometimes be just a few months, or it can be the length of a whole year. We defined rejection rate as a function of accepted, rejected, and total submissions, depending on the data that were available for each publisher. However, this definition fails to account for the dynamic status of articles as they go through peer review. For instance, publishers may define "rejection" as any article sent back to the authors, even if the result was ultimately "accept." These differences can drastically affect the absolute value of rejection rates, as some publishers may count Schrödinger-esque submissions where the underlying article is tallied as both "rejected" and "accepted" with different timestamps.

We will also note that while Frontiers and MDPI provided the underlying data used to calculate rejection rates publicly, we were forced to assemble their rejection rates manually. For Frontiers, we explicitly use *1 - (accepted articles / total submissions)* to use the same methodology as was available for MDPI data. However Frontiers reports an independent number they call "Rejected" articles that gives a different value for rejection rate when used in the formula: *rejected articles / total submissions* (Frontiers data collected from https://progressreport.frontiersin.org/peer-review, accessed Sept. 4th, 2023). However, the total "accepted" and "rejected" articles does not sum to the total submissions, indicating Frontiers is not indicating decisions on some articles that are submitted, but not yet accepted. For instance, in 2022, Frontiers reports 240,037 total submissions, 124,508 accepted, 112,104 rejected. Meanwhile, MDPI rejection rate data were available via "Journal statistics" web page html code as "total articles accepted" and "total articles rejected." We therefore defined total submissions to MDPI as the sum of all accepted and rejected articles. On the other hand, Hindawi reported their rejection rates publicly on journal pages as "acceptance rate," although the underlying calculation method is not given.

Finally, rejection rates are intrinsically tied to editorial workload capacity and total submissions received. For some journals, total workload from submissions has trade-offs with what can be feasibly edited. For instance, eLife initially saw longer times to deciding on whether to desk reject an article or not during a trial that committed to publish all articles after peer review at the author's discretion (eLife, 2019). This change to longer processing times was presumably instinctive editor behaviour to avoid the ensuing workload of accepting all articles for peer review, and so the commitment of time associated. In other journals, relatively few articles per editor might be submitted, and so more articles could be considered for publication and retained for reassessment following revisions, perhaps visible as broader turnaround time distribution curves (Fig. 3B). Thus, we will stress that the absolute rejection rate itself is not a measure of quality or rigour, but rather reflects the balance between editorial capacity, journal scope and mission, and the total submissions received.

While we could not standardise the methodology used to calculate rejection rates across publishers, we make the assumption that publishers have at least maintained a consistent methodology internally across years. For this reason, while comparing raw rejection rates comes with many caveats, comparing the direction of change itself in rejection rates shown in Fig. 4 should be relatively robust to differences between groups.

*Impact inflation*

"Impact inflation" is a new synthetic metric we define, and so here we will take care to detail its characteristics, caveats, and assumptions in depth. Principally, this metric uses the ratio of the Clarivate Impact Factor (IF) to the Scimago Journal Rank (SJR).

One of the most commonly used metrics for judging journal impact is provided by Clarivate's annual Journal Citation Reports: the journal IF. Impact Factor is calculated as the mean total citations per article in articles recently published in a journal, most commonly the last two years. The formula for IF is as follows, where $y$ represents the year of interest (Garfield, 2006):

$$IF_y = \frac{Total citations_y}{Total publications_{y-1} + Total publications_{y-2}}$$

The IF of a journal for 2022 is therefore:

$$IF_{2022} = \frac{Total citations_{2022}}{Total publications_{2021} + Total publications_{2020}}$$

The name "Impact Factor" refers to this calculation when done by Clarivate using their Web of Science database. However, the exact same calculation can be performed using other databases, including the Scopus database used by Scimago; indeed, these metrics are highly correlated (FigS14A). Because of mass delistings by Clarivate in their 2023 Journal Citation Reports that affected many journals (Quaderi, 2023), which were not delisted in Scimago data, there is a decoupling of the Scimago Cites/Doc and Clarivate IF in 2022 data (F = 6736 on 1, 3544 df, adj-$R^2$ = 0.72) compared to previous years 2012-2021 (F = 43680 on 1, 13397 df, adj-$R^2$ = 0.77). The overall trends in Impact Inflation are robust to use of either Cites/Doc or IF in 2021 or 2022 (Fig. 5A, Fig. S14). For continuity with discussion below, we will describe IF as a metric of journal "prestige," as IF is sometimes used as a proxy of journal reputation.

The Scimago Journal Rank (SJR) is a metric provided by Scimago that is calculated differently from Clarivate IF. The SJR metric is far more complex, and full details are better described in (Guerrero-Bote & Moya-Anegón, 2012). Here we will provide a summary of the key elements of the SJR that inform its relevance to Impact Inflation.

**Table 2 (repeated): Methodological differences between SJR and Impact Factor.** Table content adapted from Guerrero-Boté and Moya-Anegón (2012). Table copied here from main text for clarity of discussion.

|  | **SJR** | **Impact Factor** |
|---|---|---|
| **Database** | Scopus | Web of Science |
| **Citation time frame** | 3 years | 2 years |
| **Self-citation contribution** | Limited | Unlimited |
| **Field-weighted citation** | Weighted | Unweighted |
| **Size normalisation** | Citable document rate | Citable documents |
| **Citation networks considered?** | Yes | No |

The SJR is principally calculated using a citation network approach (visualised in Fig. S17). The reciprocal relationship of citations between journals is considered in the ultimate rank of SJR "prestige," including a higher value placed on citations between journals of the same general field. The formula used by Scimago further limits the amount of prestige that one journal can transfer to itself or to another journal. This is explicitly described as a way to avoid *"problems similar to link farms with journals with either few recent references, or too specialized"* (Guerrero-Bote & Moya-Anegón, 2012); "link farms" are akin to so-called "citation cartels" described in (Abalkina, 2023; Fister et al., 2016). This is the most important difference between IF and SJR, as IF does not consider the source of where citations come from, while SJR does. As a result, SJR does not permit journals with egregious levels of self- or co-citation to inflate the ultimate SJR prestige value.

The ratio of IF/SJR can therefore reveal journals whose total citations (IF) come from disproportionately few citing journals. MDPI journals have a much lower SJR compared to their IF. The reason for this is exemplified from the ratios of citations in/out for the flagship MDPI journals International Journal of Molecular Sciences, International Journal of Environmental Research and Public Health, and Sustainability (see Fig. S17). These three journals not only have high rates of within-journal self-citation (9.4%, 11.8%, 15.3% respectively in 2022), they and other MDPI journals further contribute the plurality of the total citations to each other (MDPI, 2021), and to other journals (e.g. Hindawi – BioMed Research International), which outside of the MDPI network are often not reciprocated (Fig. S17).

Importantly, growth of articles per journal is not an intrinsic factor behind this disparity. Frontiers has seen a similar level of growth of its articles per journal as MDPI (Fig. 1C), enabled by using the special issues model (Fig. 2), but has far lower Impact Inflation scores (Fig. 5A). Frontiers also receives more diverse citations coming from a wider pool of journals, and only sparingly from other Frontiers journals (Fig. S17). Importantly, we cannot comment on *why* these behaviours exist. What can be said is that the MDPI model of publishing seems to attract authors that cite within and across MDPI journals far more frequently than authors publishing with comparable for-profit Open Access publishers like Hindawi, Frontiers, or BioMed Central (BMC). Indeed, in a self-analysis published in 2021, MDPI's rates of within-publisher self-citation (~29%, ~500k articles) were highly elevated compared to other publishers of similar size (not an opinion shared by MDPI)(MDPI, 2021). Their rates were also higher than IEEE (~5%, ~800k articles), Wiley-Blackwell (~17%, ~1.2m articles), and Springer Nature (~24%, ~2.5m articles), lower only compared to Elsevier (~37%, ~3.1m articles) (Figure S18) (MDPI, 2021).

The ratio of IF to SJR (or of the Scimago proxy Cites/Doc to SJR) therefore assesses how two different citation-based metrics compare. The first metric (IF) is source-agnostic, counts the raw volume of citations and documents, and outputs a prestige score. The second metric has safeguards built in that prevent citation cartel-like behaviour from inflating a journal's prestige, and so if a journal receives a large number of its citations from just a few journals, it will not receive an SJR score that is proportional to its IF.

*Comment on the advertisement of IF as a metric of prestige*

It is striking to note that most journals celebrate a year-over-year increase in IF, however our study shows that IF itself has become inflated, like a depreciating currency, by the huge growth in total articles and total citations within those articles (Fig. S13, Fig. S15). As a result, unless IF is considered as a relative rank, the value of a given IF changes over time. Indeed, a publisher whose journals had an average Cites/Doc of "3.00" in 2017 was somewhat high within our publisher comparisons, however in 2022 a Cites/Doc of "3.00" is near the lowest average Cites/Doc across publishers (Fig. S11). This rapid inflation, i.e. depreciation of IF-like metrics, does not affect the underlined relative comparisons made in Clarivate Journal Citation Reports' IF rank or IF percentile. Publishers often report absolute IFs, however our analysis suggests the more accurate IF-based metric to report would be a relative rank, such as IF rank or percentile within a given category.

As our impact inflation metric is similarly proportional to IF itself, we would likewise recommend adaptations of impact inflation to compare relative ranks, such as quartiles. Unlike IF, the impact inflation metric already normalises by journal size and field by calculating SJR through citable document rate, rather than citable documents, making field-specific normalisation less important for comparisons of impact inflation across journals.

*Data and code availability*

Our analysis was based partly on publicly available data, and partly on scraped data. In both cases, our right to gather (scrape) and analyse (public, scraped) the data does not automatically translate into a right for us to *redistribute* the data. This means that we will not be able to make the data for our analysis publicly available. Nonetheless, we now provide File S2, which is a README that should guide the interested reader to assemble the publicly-available data we used the original sources. We will also provide our full analysis code upon article acceptance at: https://figshare.com/articles/figure/The_strain_on_scientific_publishing_figures_/24203790. Code and all data was made available to peer reviewers to ensure a robust peer review process.

*R packages used*

We used R version 4.3.1 (R Core Team 2023) and the following R packages: agricolae v. 1.3.6 (de Mendiburu 2023), emmeans v. 1.8.7 (Lenth 2023), ggtext v. 0.1.2 (Wilke and Wiernik 2022), gridExtra v. 2.3 (Auguie 2017), gt v. 0.9.0 (Iannone et al. 2023), gtExtras v. 0.4.5 (Mock 2022), here v. 1.0.1 (Müller 2020), hrbrthemes v. 0.8.0 (Rudis 2020), kableExtra v. 1.3.4 (Zhu 2021), magick v. 2.7.5 (Ooms 2023), MASS v. 7.3.60 (Venables and Ripley 2002), multcomp v. 1.4.25 (Hothorn, Bretz, and Westfall 2008), multcompView v. 0.1.9 (Graves, Piepho, and Sundar Dorai-Raj 2023), MuMIn v. 1.47.5 (Bartoń 2023), mvtnorm v. 1.2.2 (Genz and Bretz 2009), patchwork v. 1.1.2 (Pedersen 2022), scales v. 1.2.1 (Wickham and Seidel 2022), sjPlot v. 2.8.15 (Lüdecke 2023), survival v. 3.5.5 (Therneau T 2023), TH.data v. 1.1.2 (Hothorn 2023), tidyverse v. 2.0.0 (Wickham et al. 2019) and waffle v. 0.7.0 (Rudis and Gandy 2017).

# Package citations

## Themes in Publisher Motivation

| Publisher | Commitment to Serving Research Communities | Celebrating Growth and Size |
|---|---|---|
| Elsevier<br><br>Relx Annual report, 2023 | 'Elsevier helps researchers improve and disseminate their scientific findings through its more than 2,900 journals, enhancing the record of scientific knowledge by applying high standards of quality and ensuring trusted research can be accessed, shared and built upon. In collaboration with 33,000 editors and over 1.5m reviewers worldwide, many Elsevier journals are the foremost publications in their field, including flagship families of journals like Cell Press and The Lancet, which celebrated its 200th anniversary in 2023. Research content is distributed and accessed via ScienceDirect, the world's largest platform dedicated to peer-reviewed primary scientific and medical research.'[1]<br><br>'Researchers, especially those early in their careers and those working across disciplines face significant challenges and complexity in their daily work, including an ever-growing volume of data, prevalent misinformation and increasing workloads. Scopus AI helps them understand and explore a particular topic quickly, make connections across disciplines and collaborate with others to ensure the research has greater academic and societal impact.'[2] | 'We help ensure quality research accelerates progress for society by helping validate, improve and disseminate over 17% of the world's scientific articles. Elsevier's over 2,900 journals published more than 630,000 articles in 2023, from almost 3m submitted . . . ScienceDirect, the world's largest platform dedicated to peer-reviewed primary scientific and medical research, hosts over 21m pieces of content from over 4,700 journals and over 46,000 e-books, and has over 20m monthly unique visitors.'[3] |
| Frontiers<br><br>Annual Report, 2022 | 'Our journals now attract authors from all over the world, which means quality control is paramount. Thanks to developments in our artificial intelligence review assistant (AIRA) and the growth of our research integrity team, more than half of the articles we reject are now rejected at the desk review stage, before they enter peer review. This means that our editorial boards can focus solely on developing and publishing high quality research. | 'In 2022 we remained the sixth largest and third most-cited research publisher, with an average article citation count of 5 - a figure which beautifully captures the impact and quality of the work researchers choose to publish with us . . . We published 125,000 new research articles last year, which means we have now published more than 397,000 peer-reviewed, openly accessible articles that people all over the world can view and download openly. We also launched 49 new journals, bringing the total number of our community-led journals to 189 across more than 1,500 academic disciplines. Each |

---

[1] https://www.relx.com/~/media/Files/R/RELX-Group/documents/reports/annual-reports/relx-2023-annual-report.pdf. Page 21. Viewed 6th March 2024

[2] https://www.relx.com/~/media/Files/R/RELX-Group/documents/reports/annual-reports/relx-2023-annual-report.pdf. Page 11. Viewed 6th March 2024

[3] https://www.relx.com/~/media/Files/R/RELX-Group/documents/reports/annual-reports/relx-2023-annual-report.pdf. Page 20. Viewed 6th March 2024

| | | |
|---|---|---|
| | Researcher centricity is a core value for us at Frontiers and we value feedback from the research community as a key part of our continuous improvement. Through the course of 2022 we received more than 150,000 pieces of feedback – in the form of survey responses, emails, and in-depth interviews. We found that 92% of our authors and 88% of reviewers rate their experience as excellent or good.' [4] | journal we launch is developed by a closely-knit scientific community based on trust and integrity. More than 243,000 leading researchers choose to work on our editorial boards, pioneering in their own fields and supporting the transition to open science.' [5] |
| MDPI<br><br>Annual report, 2022 | 'The scholarly publication market is constantly changing, which creates a wealth of opportunities for innovative service providers. Our attention to authors' needs, the simplicity of our processes and the overall level of service provided continue to be the key drivers of MDPI's publication growth . . .<br><br>Increasing the visibility of scientific research is our goal. Our top five journals between them collected more than 10,000 mentions in news outlets last year. In 2023, we expect more than 100 million users on our websites. [6] | Opening paragraphs of 2022 annual report:<br><br>'2022 marked a special year for MDPI as we surpassed one million total articles published. This landmark was reached thanks to the immeasurable support, over the years, of more than 600,000 expert reviewers, 66,000 editorial board members and 6750 hard-working colleagues across MDPI's global offices.<br><br>For the fifth year in a row, MDPI was the publisher with the largest number of open access articles published. The annual growth rate exceeded 50 percent annually, on average, over the same timespan'. [7] |
| Wiley<br><br>John Wiley and Sons<br>2023 Q1 Earnings Call<br><br>'Form 10-K' 2023 Annual Report | 'The contribution of authors and their professional societies is one of the more important elements of the highly competitive publishing business. Success and continued growth depend greatly on developing new products and the means to deliver them in an environment of rapid technological change. Attracting new authors and professional societies while retaining our existing business relationships is critical to our success. If we are unable to retain our existing business relationships with authors and professional societies, this could have an adverse impact on our consolidated financial position and results of operations.' [8] | 'As of April 30, 2023, we publish over 1,900 academic research journals. We sell journal subscriptions directly to thousands of research institutions worldwide through our sales representatives, indirectly through independent subscription agents, through promotional campaigns, and through memberships in professional societies for those journals that are sponsored by societies. . . . Wiley's performance in the 2021 release of Clarivate Analytics' Journal Citation Reports (JCR) remains strong, maintaining its top 3 position in terms of citations received and sits in 4 place for journals indexed and articles published. Wiley has 7% of titles, 8% of articles, and 11% of citations.' [10] |

| | '[W]e are streamlining operations in research, particularly the referral of rejected articles from one Wiley journal to another. We've been discussing this process for some time. We call it the cascade. As a reminder, Wiley does not publish 70% of the articles we receive. Mostly this is due to improper fit between a submitted manuscript and the first journal to which it is submitted. Our goal is to find these articles another home within Wiley's portfolio, and we do this through an intelligent transfer and referral process. Notably, 65% of rejected authors are now offered another Wiley option to publish, up from 53% at year-end. It will take time to fully capitalize on this opportunity across our 1,900 journal portfolio, but we're very encouraged by the progress so far.'[9] | |
|---|---|---|
| Springer / Nature<br><br>Annual Report 2022 | 'In the research community, we build deep connections with authors and bring global visibility to their work. Through our partnerships with editors, peer reviewers and experts in our vast ecosystem, and using the latest innovations, we evaluate, quality assure, improve and publish new discoveries, many of which are free to access as we move closer to open research. This helps researchers uncover fresh ideas, advancing progress for the next generation.'[11] | 'Our research segment had another successful year, publishing record amounts of quality research, including more than 410,000 articles and more than 13,200 books. The high rates of page views, downloads and citations show these insights were widely used and shared, and that we continue to provide value for our authors.<br><br>Underlying revenue growth was 4%, outperforming the market and our key peers. This was driven by growth in the Nature Portfolio journals and in our OA business. Academic institutions continued to expand their subscriptions across the Nature Portfolio, and across all key markets. We launched new Nature titles, attracting strong levels of submissions, demonstrating the strength of the brand. Our fully OA portfolio now comprises around 600 journals, and the speed of transition to OA continues to exceed our expectations.'[12] |
| Hindawi<br><br>Wiley announces its acquisition of Hindawi for $298 million (2021) | 'The acquisition of Hindawi enables Wiley to move farther and faster toward our goal of meeting the world's urgent and escalating need for new knowledge," said Brian Napack, President and CEO, Wiley. "Hindawi is a true pioneer in the industry, empowering researchers with a fully digital, user-friendly publishing process that gets their life-changing, peer-reviewed discoveries out into the world faster and more efficiently.'[13] | 'Hindawi has a robust portfolio of over 200 peer-reviewed scientific, technical, and medical journals, a highly efficient publishing platform, and a low-cost infrastructure. Wiley's acquisition of Hindawi unlocks significant and profitable new growth by tapping deeper into the fast-growing OA market and by delivering innovative publishing services to researchers, societies, and institutions around the world. For the fiscal year ending December 31, 2020, Hindawi is |

| | | projected to generate approximately $40 million in revenue with year over year growth of 50%.' [15] |
|---|---|---|
| | 'The addition of Hindawi's journals doubles Wiley's gold (pure) OA journal portfolio and will increase author retention by giving researchers more options to publish within Wiley titles.' [14] | |
| Taylor & Francis<br><br>Informa Annual Report,t 2022 | 'We are continuing to invest in all these areas. As well as further modernising our business processes, we are continuously investing in our digital content platforms and in content discovery and data capture. This improves the speed and quality of research publication and its usability, whether through traditional editorial and peer review channels or newer open research platforms. Being flexible to customers' needs and preferences remains a priority. Our goal is to provide a full range of services through the lifecycle of a knowledge maker, as shown on the following pages, from initial study and learning products to authoring support, support when becoming a validator or reviewer, instruction and teaching services and helping to connect research output to real-world applications.'[16]<br><br>'We have partnered with over 500institutions through 23 transformative agreements globally. The positive impact has been seen across all subject areas but particularly in Social Sciences and Humanities disciplines. In the first two years of our transformative agreement in the UK, almost 80% of articles published open access through this agreement have come from Social Sciences and Humanities disciplines. These agreements have produced a model that allows researchers in all subject areas to benefit from the increased reach and impact produced by publishing open access. They also pave the way for a continued, sustainable and stable transition from publishing content funded primarily by subscriptions to the current mixed model environment, and provide a basis to continue to support customers if, in the future, | 'Our Academic Markets business, Taylor & Francis, is performing well and with consistency. Its underlying revenue growth was 3%, up from 2.4% in 2021, putting the business well on the way to meet our GAP 2 target of 4% growth by the end of 2024.<br><br>Taylor & Francis is focused on three key areas: a well-established pay-to-read business where expert research is accessed through annual or multi-year subscriptions; a fast-developing pay-to-publish and open research business where published research is supported by funding grants and made broadly available; and an advanced learning business that publishes books and ebooks in specialist subject matter categories.' [18] |
| | more want to publish in an environment that is entirely open access.'[17] | |
| PLoS | 'A focus of our work in 2022—which will continue into the future—is ensuring that we are investing in the infrastructure that enables us to deliver rigorous, open research to a diverse audience across our portfolio. Whether that means improvements in the author submission experience, adapting to new challenges in order to safeguard and strengthen research integrity, or committing to actions that further the real-world impact of the research we publish.' | 'This year we celebrated the first 1,000 papers published across five new titles launched in 2021—a significant milestone for PLOS. These new journals represent research and researchers from across the globe, and aim to forge connections between diverse disciplines and experts addressing critical issues for human health and the health of our planet. |
| | 'As the possibilities of Open Science, and expectations for researchers, change around the world, it's important that we collaborate globally to learn from and connect with all stakeholders in scholarly communication—from early career researchers to institutional leaders to policy makers. Last year we partnered with TCC Africa to work towards our shared goals of increasing awareness of the benefits of Open Science for researchers across the African continent, and influencing Open Science policy at the national, regional, and institutional level. This year, we expanded our partnerships to AAU and EASTECO. One of our first joint efforts with TCC and AAU is a series of workshops for institutional leaders focusing on increasing awareness and providing training around Open Science practices and Open Access publishing. We also established a new regional hub in Singapore which will enable us to work more closely with research stakeholders in Asia. | Financial barriers as a result of the dominant Open Access APC model remain a top concern for researchers and a focus of innovation at PLOS as we work to make Open Science publishing more equitable. In 2022—we doubled our institutional partnerships that reduce or eliminate the burden of Open Access costs for authors. We were also delighted to partner with the Einstein Foundation in Berlin to honor researchers demonstrating rigor, reliability, robustness, and transparency in their work.'[20] |
| | Perhaps our biggest achievement of the year, PLOS partnered with DataSeer to develop Open Science Indicators—a framework for measuring Open Science behaviors such as preprint posting, data-sharing, and code-sharing in a standardized way. The new framework and analysis of the data from PLOS articles as well as a sample of comparator articles will enable us to better understand the state of Open Science practice today, and identify ways in which | |

[17] https://www.informa.com/globalassets/documents/investor-relations/2023/informa-annual-report-2022.pdf page 21 viewed 7th March 2024.
[20] https://plos.org/financial-overview/ viewed 23rd May 2024

| | | |
|---|---|---|
| | we can drive measurable, meaningful change in support of best practices.<br>With grant support provided by the Wellcome Trust, we also continued two trials to increase data accessibility. We also shared our findings from the first year of our mandatory code-sharing policy at *PLOS Computational Biology* as well as the results of our investigation into code-sharing needs and behaviors more broadly. All data from our Open Science Indicators project and other research by PLOS is openly available to the public. search communication more open, effective and fair.'[19] | |
| BMC | 'With over 20 years of expertise in pioneering open access, you can trust our extensive experience to deliver high quality, impactful research and provide a supportive publishing experience for authors. If you believe, like we do, that **openness**, **transparency** and **community focus** should be at the heart of research publishing, then we would like to welcome you to the BMC family of journals.'[21] | 'At BMC we are dedicated to publishing the best open access journals across our portfolio of over 250 titles and are always striving to drive progress in biology, health sciences and medicine.'[22]<br><br>'We pride ourselves on providing a supportive and accessible service for our authors throughout the publishing process. Over one million authors have chosen to publish with us over the past 15 years because of the service and results we deliver.'[23] |

How to download datasets for total articles and researcher growth related to "The Strain on Scientific Publishing" by Hanson et al. (2023; arXiv).

## Contents

# Scimago data download and assembly instructions

1. Go to: https://www.scimagojr.com/journalrank.php
2. Download the data as per instructions in screenshot below. Note that the full dataset must be downloaded one-by-one as .csv files for each year. Place these in the folder *Data/raw_data_Scimago*

3. The script *0_Merge_raw_Scimago_data.R* will assemble all .csv files in *Data/raw_data_Scimago* into the final dataframe used in the analysis, saved as *Scimago_data_filtered.csv*.
4. As part of *Analysis.R*, there is a line in the script to run this as part of the overall script. Because this is a time-consuming data assembly. Check that a # is not hiding this script from being run. The relevant part of *Analysis.R* is:

#### 0. assemble Scimago annual .csv files into single dataframe ####

...

# source("Scripts/0_Merge_raw_Scimago_data.R")

Change the source(…) part to remove the hashtag:

source("Scripts/0_Merge_raw_Scimago_data.R")

5. Ready for Analysis.R! These data are used primarily for Figure 1, Figure 5, and related supplementary figures.

# OECD, NSF (2022), Zwetsloot et al. (2021) PhD Data

OECD data were downloaded per the following parameters in April 2023 from the link below:

https://stats.oecd.org/Index.aspx?DataSetCode=EAG_GRAD_ENTR_RATES

This is a dataframe of graduation rates at the doctoral or equivalent level, including all first-time graduates from the ISCED level less than 35 years old. An analysis of these data has already been done by others, reported here: https://master-academia.com/number-of-phds/

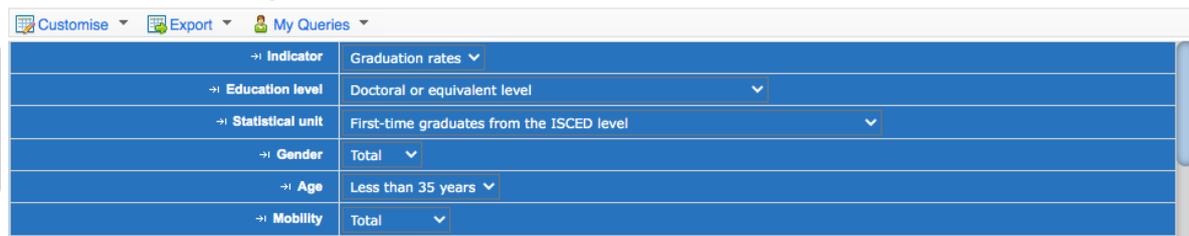

Other datasets were included to supplement numbers for India and China for consideration in the supplemental figures. Because these are amalgamations across datasets, we preferred not to present the India- and China-supplemented data in the main text, since the underlying numbers across countries were generated through different methodologies. But like this, we intended to consider whether inclusion of India and/or China might change the overall picture of global PhD growth in a meaningful way, which it did not.

NSF (2022), used for data on India. Note, China used Zwetsloot et al. (2021) because Zwetsloot et al. (2021) provided a more up-to-date dataset and included projections into future years, which meant we did not have to model this independently: https://ncses.nsf.gov/pubs/nsb20223/figure/HED-29

Zwetsloot et al. (2021), used for data on China: https://cset.georgetown.edu/wp-content/uploads/China-is-Fast-Outpacing-U.S.-STEM-PhD-Growth.pdf

These NSF and Zwetsloot et al. data were input directly into the R script ***Fig1_supp_OECD.R*** as a dataframe compiled from independent objects to make data sharing and assembly a tad easier.

# UNESCO Researchers-per-million data

Data downloaded from the UNESCO data portal:
http://data.uis.unesco.org/Index.aspx?DataSetCode=DEMO_DS

1. UNESCO researchers-per-million data were downloaded July 2023 per the settings in the screenshot below. The data were saved as ***UNESCO_researchers_per_mil.csv*** for easier processing:

2. **Some caveats to this dataset and its use:** researchers-per-million reporting is spotty or absent from year to year for some important countries. For instance, the following countries provide no data, or have absent data from 2019-on: Canada, New Zealand, Switzerland, India. From 2020-on, additionally the United States of America and China were lacking data at the time of download, and the United Kingdom lacked data but provided an estimate.

3. **Because of caveats listed in #2:** collecting data on researchers-per-million at the world level may give a skewed representation of the underlying data as this number is not informed in 2019/2020-on by countries where data are absent.